\documentclass{JHEP3}

 \def\ino{\widetilde}\def\gluino{\ino{g}}
 \def\sqk{\ino{q}}\def\squark{\ino{q}} \def\sup{\ino{u}}\def\sdn{\ino{d}}
 \def\slepton{\ino{\ell}} \def\sel{\ino{e}} \def\snu{\ino{\nu}}
 \def\gaugino{\ino{\chi}} 
 \def\M{ \overline{|\mathcal{M}|^2} }
 \def\cab{\ensuremath{C_{\alpha\beta}}}
 \def\dab{\ensuremath{\delta_{\alpha\beta}}}
 \def\prop{\ensuremath{\mathcal G}}
 \def\ckm{\ensuremath{|V_{\rm CKM}|^2}}
 \def\ckmc#1{\ensuremath{|V_{\rm CKM}[#1]|^2}}
 \def\ims #1 {\ensuremath{M^2_{[#1]}}}
 \def\zz{s-M_Z^2+iM_Z\overline\Gamma_Z}
 \def\zw{s-M_W^2+iM_W\overline\Gamma_W}
 \def\Gambr{\overline\Gamma}
   \def\IQ{{\tt IQ}}
   \def\IL{{\tt IL}}
   \def\ISQ{{\tt ISQ}}
\def\s#1{{\small#1}}

\def\HW{\s{HERWIG}}
\def\HD{\s{HDECAY}}

\def\IS{\s{ISAJET}}
\def\IW{\s{ISAWIG}}
\def\SM{\s{SM}}
\def\MSSM{\s{MSSM}}
\def\SY{\s{SUSY}}
\def\QCD{\s{QCD}}

\def\LSP{\s{LSP}}

\def\mw{M_W}
\def\mz{M_Z}

\def\as{\alpha_{\mbox{\tiny S}}}

\def\MC{Monte Carlo}

\def\qbar{\bar{q}}
\def\Qbar{\bar{Q}}
\def\dbar{\bar{d}}
\def\ubar{\bar{u}}

\def\B0bar{\overline{B^0}}

\def\l{\ell}

\def\IN{{\tt IN}}
\def\IC{{\tt IC}}

\def\al{\alpha}
\def\be{\beta}

\def\ie{{\it i.e.\/}}
\def\lam{\lambda}
\def\cht{\gaugino}
\def\upt{\sup}
\def\elt{\ino{\ell}}
\def\nut{\ino{\nu}}

 \title{Implementation of supersymmetric processes in
the HERWIG event generator
 \thanks{Work supported in part by the U.K.\ Particle Physics and
Astronomy Research Council and by the EU Fourth Framework Programme
`Training and Mobility of Researchers', Network `Quantum Chromodynamics
and the Deep Structure of Elementary Particles', contract FMRX-CT98-0194
(DG 12 - MIHT).} }

 \author{Stefano Moretti\\ Theory Division, CERN and IPPP,
         University of Durham\\ E-mail: \email{Stefano.Moretti@cern.ch}}
 \author{Kosuke Odagiri\\ Theory Group, KEK\\
         E-mail: \email{odagirik@post.kek.jp}}
 \author{Peter Richardson\\ Cavendish Laboratory and DAMTP, University
         of Cambridge\\ E-mail: \email{richardn@hep.phy.cam.ac.uk}}
 \author{Michael~H.~Seymour\\ Department of Physics \& Astronomy,
         University of Manchester\\ E-mail: \email{M.Seymour@rl.ac.uk}}
 \author{Bryan~R.~Webber\\ Cavendish Laboratory, University of
         Cambridge\\ E-mail: \email{webber@hep.phy.cam.ac.uk}}

 \abstract{
   We describe the implementation of supersymmetric processes in the \HW\
\MC\ event generator. We define relevant parameter and mixing conventions
and list the hard scattering matrix elements.
   Our implementation is based on the Minimum Supersymmetric Standard
Model, with the option of R-parity violation. The sparticle spectrum is
completely general. Both hadron--hadron and lepton--lepton collisions are
covered.
   This article supplements a separate publication in which the general
features of \mbox{\HW\ 6.2} are described, and updates the treatment of
supersymmetry to version~6.4.
 }


 \preprint{Cavendish--HEP--02/03\\
 CERN--TH/2001--177\\
 DAMTP--2002--34\\
 DCPT/01/102\\
 IPPP/01/51\\
 KEK--TH/812}

\begin{document}


 \section{Introduction}
 \label{introduction}
  \HW\footnote{The latest version of the program, together with all 
relevant information, is always available via the official \HW\ page:
{\tt http://hepwww.rl.ac.uk/theory/seymour/herwig/}.}
  \cite{HERWIG6} is a general-purpose event generator for high-energy
processes.
  A major new feature 
  is the inclusion of supersymmetric (\SY) processes, made available for
the first time in \HW\ 6.1 and extended in subsequent versions.
  This article describes the implementation, defines the parameter and
mixing conventions, and lists the hard scattering matrix elements.
  For the details of the input file format and for a more general
description of our implementation, as well as the description of the
Standard Model (\SM) physics implemented in the generator, the reader
should consult Ref.~\cite{HERWIG6}. This article supplements that paper
and extends it to version 6.4. The exact version number of the code
described in this article is 6.400.

  The key features of the \HW\ 6.4 supersymmetry implementation are as
follows:
 \begin{itemize}

 \item A general Minimal Supersymmetric Standard Model (\MSSM) particle
       spectrum.

 \item Both hadron--hadron and lepton--lepton collisions.

 \item \MSSM\ $2\to2$ sparticle production subprocesses.

 \item $2\to1$, $2\to2$ and $2\to3$ Two Higgs Doublet Model (2HDM) Higgs
       boson production subprocesses in hadron--hadron collisions, and
       $2\to2$ Higgs boson production processes in lepton--lepton
       collisions.

 \item All R-parity violating $2\to2$ scattering subprocesses in
       hadron--hadron collisions that can, but do not necessarily, have
       an $s$-channel resonance,
       and $2\to2$ scattering processes in lepton--lepton collisions, not
       necessarily with an $s$-channel resonance.

 \item Cascade decay of heavy objects according to branching fractions
       taken from an input data file, incorporating 3-body and 4-body 
       matrix elements.

 \item Colour coherence between emitted partons.

 \item Spin correlations in $2\to2$ R-parity conserving production
       processes and all cascade decay processes.

 \item Beam polarisation in lepton--lepton collisions.

 \end{itemize}

  The masses, total decay widths and branching fractions, the mixings of
gauginos, third generation sfermions and Higgs bosons, the bilinear,
trilinear and R-parity violating trilinear couplings,
  are read in from a data file. A separate program (\IW) is available
(from the \HW\ web page) to convert the data from \IS\ \cite{Baer:1987au}
and \HD\ \cite{Djouadi:1998yw} to a form suitable for \HW\ and modify it
to include the R-parity violating sector.

  In our implementation, supersymmetric particles do not radiate gluons.
This assumption is reasonable if the decay lifetimes of the coloured
sparticles (spartons) are much shorter than the \QCD\ confinement scale.
In particular, a stable gluino Lightest Supersymmetric Particle (\LSP) is
not simulated. The simulation of a long-lived light top squark would
neglect effects due to gluon emission and hadronization.

  CP-violating phases are not included for either the mixings or the
couplings.

  R-parity violating lepton--gaugino and slepton--Higgs mixing is not
considered.


 \section{Parameter and mixing conventions}
 \label{conventions}


  There are several different conventions that are commonly adopted in the
literature for the \MSSM\ Lagrangian. While most authors have chosen to
follow the conventions used in Refs.~\cite{Haber:1985rc,Gunion:1986yn},
other conventions, for example Ref.~\cite{Baer:1987au}, are also commonly
used. The conventions we have used in general follow those of
Refs.~\cite{Haber:1985rc,Gunion:1986yn}. In this section we present our
conventions for the mixing matrices for the charginos, neutralinos and
sfermions, together with the conversion between the conventions of
Refs.~\cite{Haber:1985rc,Gunion:1986yn} and Ref.~\cite{Baer:1987au}.

 \subsection{Particle content}

  The \MSSM\ particle content, expressed in terms of its mass eigenstates,
is given in Tab.~1.

  \begin{center}
  \small
  \begin{tabular}{|c|c|c|}
  \hline
  spin 0            & spin 1/2                  & spin 1
                      \rule[-0.4cm]{0cm}{1.1cm}              \\\hline\hline
  down-type squarks & down-type quarks          &                  \\
  $(\ino{d}_L,\ino{d}_R,\ino{s}_L,\ino{s}_R,\ino{b}_1,\ino{b}_2)$          &
                      $(d,s,b)$                 &                  \\\hline
  up-type squarks   & up-type quarks            &                  \\
  $(\ino{u}_L,\ino{u}_R,\ino{c}_L,\ino{c}_R,\ino{t}_1,\ino{t}_2)$          &
                      $(u,c,t)$                 &                  \\\hline
  charged sleptons  & charged leptons           &                  \\
  $(\ino{e}_L,\ino{e}_R,\ino{\mu}_L,\ino{\mu}_R,\ino{\tau}_1,\ino{\tau}_2)$&
                      $(e^-,\mu^-,\tau^-)$      &                  \\\hline
  sneutrinos        & neutrinos                 &                  \\
  $(\ino{\nu}_e,\ino{\nu}_\mu,\ino{\nu}_\tau)$                             &
                      $(\nu_e,\nu_\mu,\nu_\tau)$&\\\hline
                    & gluino                    & gluon            \\
                    & $\ino{g}$                 & $g$              \\\hline
  neutral Higgs     & neutralinos               & photon and $Z^0$ \\
  $(h^0,H^0,A^0)$   & $\ino{\chi}^0_i\ (i=1-4)$ & $(\gamma,Z^0)$   \\\hline
  charged Higgs     & charginos                 & $W^\pm$          \\
  $H^+$             & $\ino{\chi}^+_i\ (i=1-2)$ & $W^+$            \\\hline
                    & Goldstino                 &                  \\
                    & $\ino{G}$                 &                  \\\hline
  \end{tabular}
  \end{center}
  {{\bf Table 1:} Particle content of the \MSSM, expressed in terms of its
mass eigenstates. The Goldstino $\ino{G}$ is the spin-1/2 component of the
gravitino. We neglect mixing for the first two generation
sfermions.}\vspace{5mm}

  In defining our conventions we need to write the particle content in
terms of the interaction eigenstates. This is shown in Tab.~2.
Superscripts $c$ refer to charge conjugation, which, stated in the
four-component notation, is defined by:
  \begin{equation}
  \psi^c=C\bar\psi^T=i\gamma^2\gamma^0\bar\psi^T.
  \label{charge_conjugation}
  \end{equation}
  We adopt the chirality basis. The two-component case, which we refer to
when discussing mixing conventions, follows trivially.
  Acting on chirality-left and chirality-right two-component spinors, we 
have:
  \begin{eqnarray}
  (\phi_L)^c &=&  i\sigma^2\phi_L^*= \varepsilon_{ij}\phi^*_{jL},\\
  (\phi_R)^c &=& -i\sigma^2\phi_R^*=-\varepsilon_{ij}\phi^*_{jR}.
  \end{eqnarray}
  The Levi-Civita tensor $\varepsilon_{ij}$ is defined by
$\varepsilon_{12}=-\varepsilon_{21}=1,
\varepsilon_{11}=\varepsilon_{22}=0$.

 \subsection{Parameter conventions}

  In terms of the scalar fields defined in Tab.~2, we write our scalar
superpotential as follows:
  \begin{equation}\label{scalar_superpotential}
  W_s=\varepsilon_{ij}\left[
  \mu H_1^iH_2^j+f_\ell H_1^i\ino{L}^j\ino{E}+
  f_DH_1^i\ino{Q}^j\ino{D}-f_UH_2^i\ino{Q}^j\ino{U}\right].
  \end{equation}
  The above definition is consistent with
Refs.~\cite{Gunion:1986yn,Gunion:1989we}. The sign of the term
proportional to the up-type quark coupling Yukawa matrix $f_U$ is such
that the Yukawa couplings are positive.

  \begin{center}
  \small
  \begin{tabular}{|c|c|c|}
  \hline
  spin 0            & spin 1/2                  & spin 1
                      \rule[-0.4cm]{0cm}{1.1cm}              \\\hline\hline
  $\begin{array}{l}
  \ino{Q}=\left(\begin{array}{c}\ino{U}_L\\\ino{D}_L\end{array}\right)\\
  \ino{D}=\ino{D}_R^*\\
  \ino{U}=\ino{U}_R^*
  \end{array}$
  &
  $\begin{array}{l}
  Q=\left(\begin{array}{c}   u_L\\   d_L\end{array}\right)\\
  D=d_R^c\\
  U=u_R^c
  \end{array}$
  &\\\hline
  $\begin{array}{l}
  \ino{L}=\left(\begin{array}{c}\snu\\\sel_L\end{array}\right)\\
  \ino{E}=\sel_R^*
  \end{array}$
  &
  $\begin{array}{l}
  L=\left(\begin{array}{c} \nu\\   e_L\end{array}\right)\\
  E=e_R^c
  \end{array}$
  &\\\hline
  &$\ino{g}$&$g$\\\hline
  &
  $\begin{array}{l}
  \ino{B}\\
  \ino{W}=\left(\begin{array}{c}
                   \ino{W}^+\\\ino{W}_3^0\\\ino{W}^-\end{array}\right)
  \end{array}$
  &
  $\begin{array}{l}
  B\\
  W=\left(\begin{array}{c}W^+\\W_3^0\\W^-\end{array}\right)
  \end{array}$
  \\
  $         H_1=\left(\begin{array}{c}H_1^0\\H_1^-\end{array}\right)$&
  $\ino{H}_1=\left(\begin{array}{c}
                      \ino{H}_1^0\\\ino{H}_1^-\end{array}\right)$&\\
  $         H_2=\left(\begin{array}{c}H_2^+\\H_2^0\end{array}\right)$&
  $\ino{H}_2=\left(\begin{array}{c}
                      \ino{H}_2^+\\\ino{H}_2^0\end{array}\right)$&\\
  \hline
  &$\ino{G}$&\\\hline
  \end{tabular}
  \end{center}
  {{\bf Table 2:} Field content of the \MSSM, expressed in terms of its
interaction eigenstates. Generation and colour indices are suppressed.
  }\vspace{5mm}

  For the R-parity violating scalar superpotential, we take:
  \begin{equation}\label{rpvsuperpotential}
  W_{\rm{\not R_p}} = \varepsilon_{ij}\left[
  \kappa_a\ino{L}_a^{i}H^j_2+
  \frac12\lam_{abc}\ino{L}_{a}^{i}\ino{L}_{b}^{j}\ino{E}_{c}+
  \lam_{abc}'\ino{L}_{a}^{i}\ino{Q}_{b}^{j}\ino{D}_{c}\right] +
  \frac12\lam_{abc}''\varepsilon_{c_1c_2c_3}\ino{U}^{c_1}_{a}
  \ino{D}^{c_2}_{b}\ino{D}^{c_3}_{c}.
  \end{equation}
  $a,b,c$ are the generation indices. In the last term, $c_1,c_2,c_3$ are
the triplet colour indices and the totally antisymmetric tensor is defined
by $\varepsilon_{123}=1$. Summation over colour indices is implicit in the
other terms. We omit the bilinear term proportional to $\kappa_a$.

  We introduce our soft \SY-breaking bilinear and trilinear terms as
follows:
  \begin{equation}\label{soft_interactions}
  V_I=\varepsilon_{ij}\left[
  \mu BH_1^iH_2^j+f_\ell A_\tau H_1^i\ino{L}^j\ino{E}+
  f_DA_bH_1^i\ino{Q}^j\ino{D}-f_UA_tH_2^i\ino{Q}^j\ino{U}\right] +
  (\rm{h.c.}).
  \end{equation} 
  This is consistent with the definition in Ref.~\cite{Gunion:1986yn}.
  We neglect $A$ term contributions to the first two generations. The
contribution of the potential to the Lagrangian is ${\mathcal L}=-V$ in
general. (h.c.) stands for Hermitian conjugation.

  Our soft \SY-breaking masses are defined schematically as follows:
  \begin{equation}\label{soft_masses}
  V_M=\sum_\phi M^2_\phi\phi^\dagger\phi
     +\sum_\psi M_\psi\bar\psi\psi,
  \end{equation}
  where $\phi$ are the sfermion fields and $\psi$ are the gaugino and
gluino fields in the four-component notation. Summations over colour and
weak isospin indices are implicit. We write the gaugino soft-breaking
masses as $M_{\ino B}=M_1$ and $M_{\ino W}=M_2$. In the notation of
Refs.~\cite{Haber:1985rc,Gunion:1986yn}, $M'$ and $M$ are used instead of
$M_1$ and $M_2$, respectively. However, the notation $M_1$ and $M_2$ has
become more common.

  The above conventions for the signs of the $\mu$ and the $A$ terms are
the same as those adopted in \IS\ \cite{Baer:1987au} and Haber and Kane
\cite{Haber:1985rc}. This also agrees with the conventions of the other
major \SY\ event generators, \s{SPYTHIA} \cite{Mrenna:1997hu} and
\s{SUSYGEN} \cite{Katsanevas:1997fb}.
  While these conventions are the same internally, \IS\ uses a different 
notation which corresponds to the following:
  \begin{eqnarray}
  2m_1  &=& -\mu, \\
  \mu_1 &=& -M_1,\\
  \mu_2 &=& -M_2,
  \end{eqnarray}
  where $m_1$, $\mu_1$ and $\mu_2$ are in the notation of \IS.

 \subsection{Mixing conventions}

  While internally \HW\ uses the conventions of
Refs.~\cite{Haber:1985rc,Gunion:1986yn,Gunion:1989we}, the mass and decay
spectra are fed in from a data file. In particular, we have provided a
separate code (\IW) for the conversion of data from \IS\
\cite{Baer:1999sp}. This way, the masses of the sparticles and their
R-parity conserving decay rates are calculated using \IS\ and the R-parity
violating section is added by \IW. \IS\ also calculates the mixing
matrices for the electroweak gauginos and scalar fermions. \IS\ uses the
conventions of Refs.~\cite{Baer:1987au,Baer:1994xr} and it is essential
that the conversion between the two formalisms is done correctly.

  We first consider the conversion between the two formalisms for the
gauginos and then for the left--right sfermion mixing.
  We compare the Lagrangian of Ref.~\cite{Baer:1987au} against that of
Refs.~\cite{Haber:1985rc,Gunion:1986yn} to obtain the conversions between
them. We adopt the conventions of Refs.~\cite{Haber:1985rc,Gunion:1986yn}
for the chargino and neutralino mixing matrices.

 \subsubsection{Charginos}

  In the notation of Ref.~\cite{Gunion:1986yn}, we define the following
doublets of two-component spinors for the charginos in their interaction
eigenstates: 
  \begin{eqnarray}
  \psi^+ & = & \left( -i\lambda^+,\psi^+_{H_2} \right)\!,\\
  \psi^- & = & \left( -i\lambda^-,\psi^-_{H_1} \right)\!,
  \end{eqnarray}
  where $\lambda^\pm$ are the charged Winos, $\psi^-_{H_1}$ is the charged
Higgsino associated with the Higgs that gives mass to the down-type quarks
and $\psi^+_{H_2}$ is the charged Higgsino associated with the Higgs
doublet that gives mass to the up-type quarks. The mass term in the
Lagrangian from eqn.~(A.2) of Ref.~\cite{Gunion:1986yn} is:
  \begin{equation}
  \mathcal{L}_{\rm{chargino}} = -\frac12\left(\psi^+ \ \psi^-\right)
  \left(\begin{array}{cc} 0 & X^T \\ X & 0 \end{array}\right)
  \left(\begin{array}{c}  \psi^+\\ \psi^-  \end{array}\right) +
  (\rm{h.c.}),
  \end{equation}
  where the mass matrix is given by:
  \begin{equation}  
  X=\left(\begin{array}{cc} M_2 & \mw\sqrt2\sin\beta \\
           \mw\sqrt2\cos\beta & \mu \end{array}\right)\!.
  \label{eqn:charginomatrix}
  \end{equation}
  In Ref.~\cite{Gunion:1986yn} this is diagonalized in the two-component
notation by defining the 
  mass eigenstates:
  \begin{eqnarray}
  \chi^+_i &=& V_{ij}\psi^+_j,\\
  \chi^-_i &=& U_{ij}\psi^-_j,
  \end{eqnarray}
  where $U$ and $V$ are unitary matrices chosen such that 
  \begin{equation}
  U^*XV^{-1}= M_D.
  \end{equation}
  $M_D$ is the diagonal chargino mass matrix. The four-component mass
eigenstates, the charginos, are defined in terms of the two-component
fields as:
  \begin{equation}
  \ino\chi^+_1 =
  \left(\begin{array}{c}\chi^+_1\\(\chi^-_1)^c\end{array}\right)\!,
  \ \ \ \ \ \
  \ino\chi^+_2 =
  \left(\begin{array}{c}\chi^+_2\\(\chi^-_2)^c\end{array}\right)\!.
  \end{equation}
  We need to express the chargino mass terms in the Lagrangian in a
four-component notation so that we can compare this Lagrangian with the
conventions of Ref.~\cite{Baer:1987au}. We define the four-component
spinors as in Ref.~\cite{Gunion:1986yn}:
  \begin{equation}
  \ino{W}^+ = \left(\begin{array}{c}-i\lambda^+ \\ (-i\lambda^-)^c
  \end{array}\right)\!,\ \ \ \ \
  \ \ino{H}^+ = \left(\begin{array}{c}\psi^+_{H_2} \\ (\psi^-_{H_1})^c
  \end{array}\right)\!.
  \end{equation}  
  Using this notation, we can express the Lagrangian in the four-component
notation:
  \begin{equation}
  \mathcal{L}_{\rm{chargino}} = -\left(\overline{\ino{W}^+}
  \ \overline{\ino{H}^+}\right)
  \left[  XP_L + X^TP_R\right]
  \left(\begin{array}{c}  \ino{W}^+\\ \ino{H}^+
  \end{array}\right)\!.
  \label{eqn:GHchar}
  \end{equation}

  The chirality projection operators are defined as
$P_{L/R}=(1$--/$\!$+$\gamma_5)/2$.
  Eqn.~(\ref{eqn:GHchar}) has a similar form to the chargino mass
term\footnote{This Lagrangian is taken from Ref.~\cite{Tata:1997uf} which
corrects a sign error in the off-diagonal terms in
Ref.~\cite{Baer:1987au}.} in eqn.~(2.1) of Ref.~\cite{Baer:1987au}:
  \begin{equation}
  \mathcal{L}_{\rm{chargino}} =
  -\left(\bar{\lambda}\ \ \bar{\chi}\right)
   \left[M_{\rm{charge}}P_L+M^T_{\rm{charge}}P_R\right]
   \left(\begin{array}{c} \lambda\\\chi\end{array}\right)\!,
  \end{equation}
 where
  \begin{equation}  
  M_{\rm{charge}} = \left(\begin{array}{cc} \mu_2 & -gv' \\ -gv &
  2m_1\end{array}\right)
  = -\left(\begin{array}{cc} M_2 & \sqrt2\mw\cos\beta \\
  \sqrt2\mw\sin\beta  & \mu\end{array}  \right)\!,
  \end{equation}
  where $v$ and $v'$ are the vacuum expectation values for the Higgs
fields that give mass to the up- and down-type quarks, respectively.
  We now come to the problem of comparing the notation of
Ref.~\cite{Gunion:1986yn} against that of Ref.~\cite{Baer:1987au}. The
Wino and Higgsino fields of Ref.~\cite{Baer:1987au} are the charge
conjugates of those used in Ref.~\cite{Gunion:1986yn}, so that the
following transformation is required:
  \begin{eqnarray}
  \ino{W}^+ &=& \lambda^c,\\
  \ino{H}^+ &=& \chi^c.
  \label{eqn:Cfields}
  \end{eqnarray}
  This gives the chargino mass matrix in the form:
  \begin{equation}
  \mathcal{L}_{\rm{chargino}} =
  \left(\overline{\ino{W}^+},\overline{\ino{H}^+}\right)
  \left[M'_{\rm{charge}}P_L+{M'}^T_{\rm{charge}}P_R\right]
  \left(\begin{array}{c}\ino{W}^+\\\ino{H}^+\end{array}\right)\!,
  \end{equation}  
  where
  \begin{equation}   
  M'_{\rm{charge}} = \left(\begin{array}{cc} -\mu_2 & gv \\
  gv' & -2m_1\end{array}\right)
   = \left(\begin{array}{cc} M_2 & \sqrt2\mw\sin\beta \\
  \sqrt2\mw\cos\beta & \mu\end{array}
  \right)\!.
  \end{equation}  
  This agrees with eqn.~(\ref{eqn:GHchar}) apart from the overall sign. We
need to express the fields in terms of the \IS\ mixing matrices. The
\IS\ mixing matrices are given in eqns.~(2.10) and (2.11) of
Ref.~\cite{Baer:1987au} as:
  \begin{eqnarray}
  \left(\begin{array}{c} \ino{W}_+ \\
  \ino{W}_-\end{array}\right)_L & = &
  \left( \begin{array}{cc}\theta_x\cos\gamma_L &
  -\theta_x\sin\gamma_L \\ \phantom{\theta_x}\sin\gamma_L &
  \phantom{-\theta_x}\cos\gamma_L \end{array}\right)
  \left(\begin{array}{c} \lambda\\ \chi \end{array}\right)_L,\\[0.5mm]
  \left(\begin{array}{c} (-1)^{\theta_+}\ino{W}_+ \\
  (-1)^{\theta_-}\ino{W}_-\end{array}\right)_R
  & = &
  \left( \begin{array}{cc}\theta_y\cos\gamma_R & -\theta_y\sin\gamma_R \\
  \phantom{\theta_y} \sin\gamma_R &
  \phantom{-\theta_y}\cos\gamma_R \end{array}\right)
  \left(\begin{array}{c} \lambda \\ \chi \end{array}\right)_R.
  \end{eqnarray}  
  The mixing angles $\gamma_L$ and $\gamma_R$, and the sign functions
$\theta_x$, $\theta_y$, $\theta_+$, and $\theta_-$ are defined in
Ref.~\cite{Baer:1987au}.

  We can transform these equations into the notation of
Ref.~\cite{Haber:1985rc} with the identification:
  \begin{eqnarray}
  \ino\chi_1^+ &=& \ino{W}_-^c,\\
  \ino\chi_2^+ &=& \ino{W}_+^c.
  \end{eqnarray}
  This gives:
  \begin{eqnarray}
  P_L\left(\begin{array}{c} \ino{W}^+ \\ \ino{H}^+\end{array}\right) &=&
  P_L\left(\begin{array}{cc} -\sin\gamma_R & -\theta_y\cos\gamma_R \\
                             -\cos\gamma_R &  \theta_y\sin\gamma_R
  \end{array}\right)
  \left(\begin{array}{c} (-1)^{\theta_-}\ino\chi_1^+ \\
  (-1)^{\theta_+}\ino\chi_2^+\end{array}
  \right)\!,\\[0.5mm]
  P_R\left(\begin{array}{c} \ino{W}^+ \\
  \ino{H}^+\end{array}\right) &=&
  P_R\left(\begin{array}{cc} -\sin\gamma_L & -\theta_x\cos\gamma_L \\
                             -\cos\gamma_L &  \theta_x\sin\gamma_L
  \end{array}\right)
  \left(\begin{array}{c} \ino\chi_1^+ \\ \ino\chi_2^+\end{array}\right)\!.
  \label{eqn:ISAJETcharginomix}
  \end{eqnarray}
  These mixing matrices are now in a form that allows us to compare them
with the notation of Ref.~\cite{Gunion:1986yn}, eqn.~(A.13):
  \begin{eqnarray}
  P_L\left(\begin{array}{c} \ino{W}^+ \\
  \ino{H}^+\end{array}\right) &=&
  P_L\left(\begin{array}{cc} V^*_{11} & V^*_{21} \\
                             V^*_{12} & V^*_{22}\end{array}\right)
  \left(\begin{array}{c} \ino\chi_1^+ \\ \ino\chi_2^+\end{array}\right)\!,
  \\[0.5mm] P_R\left(\begin{array}{c} \ino{W}^+ \\
  \ino{H}^+\end{array}\right) &=&
  P_R\left(\begin{array}{cc} U_{11} & U_{21} \\ U_{12} & U_{22}
  \end{array}\right)
  \left(\begin{array}{c} \ino\chi_1^+ \\ \ino\chi_2^+\end{array}\right)\!.
  \label{eqn:HKcharginomix}
  \end{eqnarray}
  By comparing eqns.~(\ref{eqn:ISAJETcharginomix}) and
(\ref{eqn:HKcharginomix}) we obtain the mixing matrices in the notation
of Ref.~\cite{Gunion:1986yn}:
  \begin{eqnarray}
  U &=& \left(\begin{array}{cc} \phantom{\theta_x}-\sin\gamma_L &
  \phantom{\theta_x}-\cos\gamma_L \\  -\theta_x\cos\gamma_L &
  \phantom{-}\theta_x\sin\gamma_L
  \end{array}\right)\!,\\[0.5mm]
  V &=& \left(\begin{array}{cc} \phantom{\theta_y}-\sin\gamma_R
  (-1)^{\theta_-} & \phantom{\theta_y}-\cos\gamma_R
  (-1)^{\theta_+} \\ -\theta_y\cos\gamma_R (-1)^{\theta_-} &
  \phantom{-}\theta_y\sin\gamma_R (-1)^{\theta_+}
  \end{array}\right)\!.
  \end{eqnarray}
  It should be noted that we adopt the opposite sign convention for the
chargino masses due to the sign differences in the two Lagrangians.

 \subsubsection{Neutralinos}

  We define a quadruplet of two-component fermion fields for the
neutralinos in the interaction eigenstate as:
  \begin{equation}
  \psi^0 =
  \left(-i\lambda',-i\lambda^3,\psi^0_{H_1},\psi^0_{H_2}\right),
  \end{equation}
  where $\lambda'$ is the Bino, $\lambda^3$ is the neutral Wino,
$\psi^0_{H_1}$ is the Higgsino corresponding to the neutral component of
the Higgs doublet giving mass to the down-type quarks and $\psi^0_{H_2}$
is the Higgsino corresponding to the neutral component of the Higgs
doublet giving mass to the up-type quarks. The Lagrangian for the
neutralino masses from equation (A.18) of Ref.~\cite{Gunion:1986yn} is:
  \begin{equation}
  \mathcal{L}_{\rm{neutralino}} =
  -\frac12\left(\psi^0\right)^TY\psi^0+(\rm{h.c.}),
  \end{equation}
  where
  \begin{equation}
  \renewcommand{\arraycolsep}{4pt}
  Y = \left(\begin{array}{cccc}
  M_1 & 0 &-\mz\sin\theta_W\cos\beta&\phantom{-}\mz\sin\theta_W\sin\beta\\
  0 & M_2 &\phantom{-}\mz\cos\theta_W\cos\beta&-\mz\cos\theta_W\sin\beta\\
  -\mz\sin\theta_W\cos\beta &\phantom{-}\mz\cos\theta_W\cos\beta&0&-\mu\\
  \phantom{-}\mz\sin\theta_W\sin\beta &-\mz\cos\theta_W\sin\beta &-\mu 
&0\\
  \end{array}\right)\!.
  \label{eqn:neutralinomatrix}
  \end{equation}
  In Ref.~\cite{Gunion:1986yn}, the Lagrangian is diagonalized in this
two-component notation to give the mass eigenstates. The diagonalization
is performed by defining two-component fields:
  \begin{equation}  \chi^0_i = N_{ij}\psi^0_j,\qquad i,j=1,\ldots\,\!,4,
  \end{equation}
  where $N$ is a unitary matrix satisfying $N^*\, Y\, N^{-1} = N_D$, and
$N_D$ is the diagonal neutralino mass matrix. The four-component mass
eigenstates, the neutralinos, can be defined in terms of the two-component
fields:
  \begin{equation}
  \ino\chi^0_i=\left(\begin{array}{c}\chi^0_i \\
  (\chi^0_i)^c\end{array}\right)\!.
\end{equation}
  Rather than adopting this approach we need to express the neutralino
mass terms in the four-component notation before performing the
diagonalization in order to compare this Lagrangian with that of
Ref.~\cite{Baer:1987au}. We use the standard procedure of
Ref.~\cite{Gunion:1986yn} to express this in the four-component notation
by defining four-component Majorana fields:
  \begin{equation}
  \ino{B}   = \left(\begin{array}{c}-i\lambda'  \\
              (-i\lambda')^c\end{array}\right)\!,\ \ \ \ \
  \ino{W}_3 = \left(\begin{array}{c}-i\lambda^3 \\
              (-i\lambda^3)^c\end{array}\right)\!,\ \ \ \ \
  \ino{H}_1 = \left(\begin{array}{c}\psi^0_{H_1} \\
              (\psi^0_{H_1})^c\end{array}\right)\!,\ \ \ \ \
  \ino{H}_2 = \left(\begin{array}{c}\psi^0_{H_2} \\
              (\psi^0_{H_2})^c\end{array}\right)\!.
  \end{equation}
  As defined in Tab.~2, $\ino{B}$ is the Bino field, $\ino{W}_3$ is the
neutral Wino field, $\ino{H}_1$ is the field for the Higgsino associated
with the neutral component of the Higgs doublet that gives mass to the
down-type quarks and $\widetilde{H}_2$ is the field for the Higgsino
associated with the neutral component of the Higgs doublet that gives mass
to the up-type quarks.

  This gives the Lagrangian in the four-component notation:
  \begin{equation}
  \mathcal{L}_{\rm{neutralino}} =-\frac12
  \left(\overline{\ino{B}}\ \overline{\ino{W}_3}\
  \overline{\ino{H}_1},\overline{\ino{H}_2}\right)
  \left[Y P_L+Y P_R\right]\left(\begin{array}{c}
  \ino{B} \\\ino{W}_3\\\ino{H}_1\\ \ino{H}_2
  \end{array}\right)\!.
  \label{eqn:GHneut}
  \end{equation}  
  We can now compare this with the relevant Lagrangian, in the
four-component notation, given in Ref.~\cite{Baer:1987au}. This
Lagrangian\footnote{This Lagrangian is taken from Ref.~\cite{Tata:1997uf}
which corrects a sign error in Ref.~\cite{Baer:1987au}.} is from
Ref.~\cite{Baer:1987au}, eqn.~(2.2):
  \begin{equation}
  \mathcal{L}_{\rm{neutralino}}=-\frac12
  \left( \bar{h}^0,\bar{h}^{'0} , \bar{\lambda}_3, \bar{\lambda_0} \right)
  \left[ M_{\rm{neutral}}P_L+M_{\rm{neutral}}P_R\right]
  \left(\begin{array}{c} 
  h^0\\ {h'}^0\\ \lambda_3\\ \lambda_0
  \end{array}\right)\!.
  \end{equation}
  $h^0$ is the Higgsino partner of the neutral component of the Higgs
doublet that gives mass to the up-type quarks, ${h'}^0$ is the Higgsino
partner of the Higgs boson that gives mass to the down-type quarks,
$\lambda_3$ is the neutral Wino and $\lambda_0$ is the Bino. The mass 
matrix
is given by:
 \begin{equation} {\renewcommand{\arraycolsep}{10pt}
 \renewcommand{\arraystretch}{1.5}
 M_{\rm{neutral}} = \left(\begin{array}{cccc}
  0      & -2m_1         & - gv/\sqrt2         & \phantom{-} g'v/\sqrt2 \\
  -2m_1  & 0             & \phantom{-} gv'/\sqrt2  & - g'v'/\sqrt2 \\
  - gv/\sqrt2            & \phantom{-} gv'/\sqrt2  & \mu_2        & 0 \\
  \phantom{-} g'v/\sqrt2 & -g'v'/\sqrt2            & 0            & \mu_1
 \end{array}\right)\!.
 \label{eqn:ISAJETneutmass}}
 \end{equation}

  It is easier to compare this with eqn.~(\ref{eqn:GHneut}) after
reordering the entries and reexpressing it in terms of $\mz$, $\beta$ and
$\theta_W$. This gives:
  \begin{equation}
  \mathcal{L}_{\rm{neutralino}}=\frac12
  \left( \bar{\lambda_0}, \bar{\lambda}_3, \bar{h}^{'0},\bar{h}^0  \right)
  \left[ M'_{\rm{neutral}}P_L+M'_{\rm{neutral}}P_R\right]
  \left(\begin{array}{c}
  \lambda_0 \\ \lambda_3\\{h'}^0\\h^0
  \end{array}\right)\!,
  \end{equation}
  where:
  \begin{equation}
  {\renewcommand{\arraycolsep}{3pt}
  M'_{\rm{neutral}} = \left(\!\!\begin{array}{cccc}
  M_1&0&\phantom{-}\mz\sin\theta_W\cos\beta&-\mz\sin\theta_W\sin\beta \\
  0&M_2&-\mz\cos\theta_W\cos\beta&\phantom{-}\mz\cos\theta_W\sin\beta \\
  \phantom{-}\mz\sin\theta_W\cos\beta&-\mz\cos\theta_W\cos\beta&0&-\mu \\
  -\mz\sin\theta_W\sin\beta&\phantom{-}\mz\cos\theta_W\sin\beta&-\mu&0 \\
  \end{array}\right)\!.}
  \end{equation}
  This equation then agrees with eqn.~(\ref{eqn:GHneut}) up to an overall
sign, provided we make the following identification:
  \begin{eqnarray}
  \widetilde{B} &  = &  \lambda_0,  \\
  \widetilde{W}_3 &=& \lambda_3,\\
  \widetilde{H}_1 &=&-{h'}^0,\\  
  \widetilde{H}_2 &=&-h^0.
  \end{eqnarray}
  The convention from \IS\ for the mixing,
taken from eqn.~(2.12) of Ref.~\cite{Baer:1987au}, is
  \begin{equation}
  \left(\begin{array}{c}\left(-i\gamma_5\right)^{\theta_1}\ino{Z}_1\\
                        \left(-i\gamma_5\right)^{\theta_2}\ino{Z}_2\\
                        \left(-i\gamma_5\right)^{\theta_3}\ino{Z}_3\\
        \left(-i\gamma_5\right)^{\theta_4}\ino{Z}_4\end{array}\right)
  =  \left(\begin{array}{cccc}
       v_1^{(1)} & v_2^{(1)} & v_3^{(1)} & v_4^{(1)} \\
       v_1^{(2)} & v_2^{(2)} & v_3^{(2)} & v_4^{(2)} \\
       v_1^{(3)} & v_2^{(3)} & v_3^{(3)} & v_4^{(3)} \\
       v_1^{(4)} & v_2^{(4)} & v_3^{(4)} & v_4^{(4)}
     \end{array}\right)
  \left(\begin{array}{c}h^0\\{h'}^0\\\lambda_3\\\lambda_0
  \end{array}\right)\!,
  \end{equation}
  where $\ino{Z}_i$ are the mass eigenstates obtained by diagonalizing the
mass matrix given in eqn.~(\ref{eqn:ISAJETneutmass}), $v_j^{(i)}$ are the
elements of the mixing matrix, and $\theta_i$ is zero (one) if the mass of
$\ino{Z}_i$ is positive (negative).

  After reordering, we get:
  \begin{equation}
  \left(\begin{array}{c}\left(-i\gamma_5\right)^{\theta_1}\ino{Z}_1\\
                        \left(-i\gamma_5\right)^{\theta_2}\ino{Z}_2\\
                        \left(-i\gamma_5\right)^{\theta_3}\ino{Z}_3\\
        \left(-i\gamma_5\right)^{\theta_4}\ino{Z}_4\end{array}\right)
  =  \left(\begin{array}{cccc}
       v_4^{(1)} & v_3^{(1)} & -v_2^{(1)} & -v_1^{(1)} \\
       v_4^{(2)} & v_3^{(2)} & -v_2^{(2)} & -v_1^{(2)} \\
       v_4^{(3)} & v_3^{(3)} & -v_2^{(3)} & -v_1^{(3)} \\
       v_4^{(4)} & v_3^{(4)} & -v_2^{(4)} & -v_1^{(4)}
     \end{array}\right)
  \left(\begin{array}{c} \ino{B} \\ \ino{W} \\ \ino{H}_1  \\ \ino{H}_2
  \end{array}\right)\!.
  \end{equation}
  We can therefore obtain the mixing matrix in the notation of
Ref.~\cite{Gunion:1986yn} by making the identification:
  \begin{eqnarray}
        N_{i1}& =& v_4^{(i)}, \\
        N_{i2}& =& v_3^{(i)},\\
        N_{i3}& =& -v_2^{(i)}, \\
        N_{i4}& =& -v_1^{(i)}.
  \end{eqnarray}
  Again, we need to adopt the opposite sign convention for the neutralino
masses.

\subsubsection{Left--right sfermion mixing}


  In addition to the mixing of the neutralinos and charginos, we need to
consider the left--right mixing of the sfermions. In general, as the
off-diagonal terms in the mass matrices are proportional to the fermion
mass, these effects are only important for the third generation sfermions,
\ie\ stop, sbottom and stau. We describe our convention for the top
squarks. The others are treated analogously.

  The following mass matrix for the top squarks uses the
conventions\footnote{This is the same as Ref.~\cite{Bartl:1994bu}, SPYTHIA
\cite{Mrenna:1997hu} and SUSYGEN \cite{Katsanevas:1997fb}.} of
Ref.~\cite{Gunion:1986yn} and is taken from eqn.~(4.17) therein.
  \begin{equation}
  M^2_{\ino{t}} = \left(\begin{array}{cc}
  M^2_{\ino{Q}}+M^2_Z\cos2\beta
  \left(\frac12-\frac23\sin^2\theta_W\right)+m^2_t &
  m_t\left(A_t-\mu\cot\beta\right) \\
  m_t\left(A_t-\mu\cot\beta\right) &
  M^2_{\ino{U}}+\frac23m^2_Z\cos2\beta\sin^2\theta_W+m^2_t
  \end{array}\right)\!,
  \end{equation}
  where $M_{\ino{Q}}$ and $M_{\ino{U}}$ are soft \SY-breaking masses for
the left and right top squarks, respectively. $A_t$ is the trilinear soft
\SY-breaking term for the interaction of the left and right stop squarks
with the Higgs boson. This compares with the \IS\ matrix from
Ref.~\cite{Baer:1994xr}:
  \begin{equation}
  M^2_{\ino{t}} = \left(\begin{array}{cc}
  M^2_{\ino{t}_L}+M^2_Z\cos2\beta
  \left(\frac12-\frac23\sin^2\theta_W\right)+m^2_t &
  -m_t\left(A_t-\mu\cot\beta\right) \\
  -m_t\left(A_t-\mu\cot\beta\right) &
  M^2_{\ino{t}_R}+\frac23m^2_Z\cos2\beta\sin^2\theta_W+m^2_t
  \end{array}\right)\!,
\end{equation}
  where $M^2_{\ino{t}_L}=M^2_{\ino{Q}}$ and
$M^2_{\ino{t}_R}=M^2_{\ino{U}}$. There is a difference in the sign of the
off-diagonal terms. This means that, as the sign conventions for the $\mu$
and $A$ terms are the same, there is a difference in the relative phases
of the two fields in the different conventions. Hence we should apply the
following change to the \IS\ output:
\begin{equation}
\theta_t \longrightarrow -\theta_t,
\end{equation}
 \ie\ change the sign of the stop mixing angle. The same argument also
applies to the sbottom and stau mixing angles.
 
  We adopt the following convention for the sfermion mixing matrices:
  \begin{equation}
  \left(\begin{array}{c}\ino{q}_{iL}\\\ino{q}_{iR}\end{array}\right)
  \quad = \quad \left(\begin{array}{cc}\phantom{-}\cos\theta^i_q&
                           \phantom{-}\sin\theta^i_q \\
         -\sin\theta^i_q & \phantom{-}\sin\theta^i_q\end{array}\right)
  \left(\begin{array}{c}\ino{q}_{i1} \\ \ino{q}_{i2}\end{array}\right)
  \quad = \quad \left(\begin{array}{cc}Q^i_{L1}&Q^i_{L2}\\
                           Q^i_{R1}&Q^i_{R2}\end{array}\right)
  \left(\begin{array}{c}\ino{q}_{i1} \\ \ino{q}_{i2}\end{array}\right)\!,
  \end{equation}  
  where $\ino{q}_{iL}$ and $\ino{q}_{iR}$ are the left and right squark
fields, for the $i$th quark, where $i$ runs over quark flavours $d, u, s,
c, b, t$. $\ino{q}_{i1}$ and $\ino{q}_{i2}$ are the squark fields for the
mass eigenstates, for the quark $i$, and $\theta^i_q$ is the mixing angle
obtained by diagonalizing the mass matrix.

  Similarly, we denote the slepton mixing as above with the matrix
$L^i_{\alpha\beta}$ where $i$ runs over the charged and neutral lepton
flavour. $\beta$ is the mass eigenstate and $\alpha$ is the left/right
eigenstate. As we do not include the right-handed neutrino we omit the
left--right mixing for the sneutrinos. Note that our convention for the
mixing matrix is opposite to that for the gauginos, in the sense that here
the mixing matrices $Q^i$ act on mass eigenstates to give the interaction
eigenstates.

\subsubsection{Higgs bosons}

  The Higgs doublets are diagonalised as follows. As in Tabs.~1 and 2,
$H_i$ denote the $SU(2)$ doublet interaction eigenstates which form the
physical eigenstates $h^0, H^0, A^0, H^\pm$ after the electroweak symmetry
breaking.
  In the charged sector:
  \begin{eqnarray}
  H_2^+  &=&H^+\cos\beta+G^+\sin\beta,\nonumber\\
  H_1^-  &=&H^-\sin\beta-G^-\cos\beta,
  \label{physicalhiggsescharged}\end{eqnarray}
  where $G^\pm$ are the Goldstone modes. The neutral Higgs sector is
parametrised in the interaction eigenstate as follows:
  \begin{eqnarray}
  &&H_1^0=\frac{v\cos\beta+\eta_1+i\zeta_1}{\sqrt{2}},\nonumber\\
  &&H_2^0=\frac{v\sin\beta+\eta_2+i\zeta_2}{\sqrt{2}}.
  \end{eqnarray}
  $v$ is the vacuum expectation value.
  The diagonalisation is achieved by:
  \begin{eqnarray}
  \zeta_2&=&A^0\cos\beta+G^0\sin\beta,\nonumber\\
  \zeta_1&=&A^0\sin\beta-G^0\cos\beta,\nonumber\\
  \eta_2&=&H^0\sin\alpha+h^0\cos\alpha,\nonumber\\
  \eta_1&=&H^0\cos\alpha-h^0\sin\alpha.
  \label{physicalhiggsesneutral}\end{eqnarray}
  Our notation is consistent with Ref.~\cite{Gunion:1986yn}, but the sign
of $\alpha$ is opposite to that of \IS\ \cite{Baer:1987au}.


 \section{Matrix elements}
 \label{me}

 \subsection{Notation}\label{notation}
  For $2\to2$ hard scattering subprocesses of the form $12\to 34$, let us
define the usual Mandelstam variables in terms of the partonic momenta as
$\hat s=(p_1+p_2)^2$, $\hat t=(p_1-p_3)^2$ and $\hat u=(p_1-p_4)^2$.  For
simplicity, the `hat' will be omitted in the following.
  As initial state partons are taken to have zero kinematic mass, it
follows that $s+t+u=m_3^2+m_4^2$, $s+t_3+u_4=0$, and
$ut-m_3^2m_4^2=sp_T^2$ where $t_3=t-m_3^2$, $u_4=u-m_4^2$, and $p_T$ is
the outgoing transverse momentum.

  The electromagnetic coupling is $e$, such that $\alpha_{\rm EM}=
e^2/4\pi$. The strong coupling is $g_s$, such that $\alpha_s=g_s^2/4\pi$.
The \QCD\ colour factors are $N_C = 3$ and $C_F = (N_C^2-1)/2N_C = 4/3$.

  Colour flow directions are not stated explicitly as they are trivial in
most of the processes whose matrix elements are given below. For a general
treatment, see Ref.~\cite{Odagiri:1998ep}.
  Matrix elements given here are summed over the final state colour and
spin, and averaged over the initial state colour and spin.
  As usual, this is indicated by a long overline.
  Statistical factors of two for identical neutralinos in the final state
are written explicitly.

  Let us now consider the widths $\Gamma(q^2)$ appearing in the
propagators of unstable massive particles with mass $M(q^2)$. In order to
simplify the notation, let us define a function $\Gambr(q^2)$, which is
defined for positive $q^2$ by:
  \begin{equation}
   \sqrt{q^2} \Gamma(q^2) \equiv M(q^2) \bar\Gamma(q^2).
  \end{equation}
  For negative $q^2$, the width is taken to be zero.
  The propagator is then given in general, again for positive $q^2$, by:
  \begin{equation}
   \frac1{q^2-M^2(q^2) + i \sqrt{q^2}\Gamma(q^2)}\equiv
   \frac1{q^2-M^2(q^2) + i M(q^2) \Gambr(q^2)}.
  \end{equation}
  In the \HW\ implementation, we adopt the naive running width 
approximation and neglect the running of mass, so that:
  \begin{equation}
   \bar\Gamma(q^2) = q^2 \Gamma(M^2)/M^2.
  \end{equation}

  The following coupling notations are used in processes involving
electroweak interactions:
  \begin{eqnarray}
   Q_f   &=& U(1)_{\rm EM} {\rm \ charge}; \\
   I_f^3 &=& {\rm weak\ isospin};        \\
   L_f   &=& (I_f^3 - Q_f\sin^2\theta_W)/\cos\theta_W\sin\theta_W; 
                            \label{def_Lf} \\
   R_f   &=& -Q_f\sin^2\theta_W/\cos\theta_W\sin\theta_W.
                            \label{def_Rf}
  \end{eqnarray}
  $Q_f$ and $I_f^3$ are as listed in Tab.~3.
  In terms of the above, we define the neutralino to sfermion coupling
coefficients as:
  \begin{eqnarray}
   \ino{L}_{fa} &=&
      \left( N_{a1}\cos\theta_W+N_{a2}\sin\theta_W\right)Q_f+
      \left(-N_{a1}\sin\theta_W+N_{a2}\cos\theta_W\right)L_f; \\
   \ino{R}_{fa} &=&
      \left( N_{a1}\cos\theta_W+N_{a2}\sin\theta_W\right)Q_f+
      \left(-N_{a1}\sin\theta_W+N_{a2}\cos\theta_W\right)R_f.
  \end{eqnarray}
  Following Ref.~\cite{Haber:1985rc},
  we have neglected the contribution of the Higgsino component, which is
proportional to the Yukawa coupling. This affects the cross sections for
sfermion--Higgsino associated production, namely the Higgsino component of
the processes $gb\to\ino{t}\ino{\chi}^\pm,\ino{b}\ino{\chi}^0$.

  \begin{center}
  \small
  \begin{tabular}{|c|c|c|}
  \hline
  $f$   & $Q_f$  & $I_f^3$ \\\hline
  $e$   & $-1$   & $-1/2$  \\
  $\nu$ & $0$    & $+1/2$  \\
  $d$   & $-1/3$ & $-1/2$  \\
  $u$   & $+2/3$ & $+1/2$  \\\hline
  \end{tabular}
  \end{center}
  {{\bf Table 3:} The $U(1)_{\rm EM}$ charge $Q_f$ and the weak isospin
$I_f^3$ for the Standard Model fermions.
  }\vspace{5mm}

 \subsection{Sparton pair production}
  The matrix elements and the colour decompositions for $2\to2$
parton-to-sparton scattering are given in Ref.~\cite{Odagiri:1998ep}. The
term `sparton' refers to coloured superparticles, namely the squarks and
the gluino. Neglecting the squark left--right mixings, the
lepton-annihilation reaction $\ell^-\ell^+\to \squark\squark^*$ is given
by:
  \begin{eqnarray}
   \M_{(L,R)}&=&e^4N_Csp_T^2 \nonumber\\ &\times& \left[
   \left| \frac{Q_\ell Q_q}{s}+\frac{L_\ell(L_q,R_q)}{\zz} \right|^2
  +\left| \frac{Q_\ell Q_q}{s}+\frac{R_\ell(L_q,R_q)}{\zz} \right|^2
   \right].\label{ee_spartons}
  \end{eqnarray}
  The generalization to the case with left--right mixings is given by:
  \begin{equation}
   (L_q,R_q) \to L_qQ_{Li}Q_{Lj}+
                 R_qQ_{Ri}Q_{Rj}.
   \label{ee_spartons_mixed}
  \end{equation}
  The indices $i,j$ label the squarks that are produced.

 \subsection{Slepton pair production}
  The matrix elements are similar to eqns.~(\ref{ee_spartons}) and
(\ref{ee_spartons_mixed}). For the quark annihilation reaction $q\bar
q\to\slepton\slepton^*$ one should interchange the indices
$(q\leftrightarrow\ell)$, and change the colour factor from $N_C$ to
$1/N_C$. In addition, for the charged current reaction $q\bar
q'\to\slepton_L\ino\nu^*$ we have
  \begin{equation}
   \M(q\bar q'\to\slepton_L\ino\nu^*)=
   \frac{e^4sp_T^2}{\left(2\sin^2\theta_W\right)^2N_C}\,
   \left|\frac1\zw\right|^2.
  \end{equation}
  For the lepton
annihilation reaction the colour factor is $1$, and there is an extra
$t$-channel gaugino contribution to the production of slepton pairs of
the same generation as the colliding lepton pair, namely,
  \begin{eqnarray}
   \M(e^-e^+\to\ino{e}_L\ino{e}_L^*,\ino{e}_R\ino{e}_R^*)
   &=&e^4sp_T^2 \nonumber\\ &\times& \Biggl[
   \left| \frac{1}{s}+\frac{L_e(L_e,R_e)}{\zz}+\sum_{a=1}^4
   \left(\frac{\ino{L}_{ea}^2}{t-M^2_{\ino\chi^0_a}},0\right)\right|^2
   \nonumber\\ &+&
   \left| \frac{1}{s}+\frac{R_e(L_e,R_e)}{\zz}+\sum_{a=1}^4
   \left(0,\frac{\ino{R}_{ea}^2}{t-M^2_{\ino\chi^0_a}}\right)\right|^2
   \Biggr],
   \label{ee_sleptons_llrr}\\
   \M(e^-e^+\to\ino{e}_L\ino{e}_R^*,\ino{e}_R\ino{e}_L^*)&=&e^4
   \left(\sum_{a=1}^4
   \frac{M_{\ino\chi^0_a}\sqrt{s}}{t-M^2_{\ino\chi^0_a}}
   \ino{L}_{ea}\ino{R}_{ea}\right)^2.
   \label{ee_sleptons_lrrl}
  \end{eqnarray}
  The neutralino masses appearing above are the signed masses. The
production of sneutrinos of the same generation as the colliding lepton
pair is given by:
  \begin{eqnarray}
   \M(e^-e^+\to\ino{\nu}_{eL}\ino{\nu}_{eL}^*) &=& e^4sp_T^2\nonumber\\
    &\times& \Biggl[
   \left| \frac{L_eL_\nu}{\zz}+\sum_{a=1}^2
    \frac{(V_{a1}^{\ino\chi^\pm})^2/2\sin^2\theta_W}
    {t-M^2_{\ino\chi^\pm_a}}\right|^2 \nonumber\\&+&
   \left| \frac{R_eL_\nu}{\zz} \right|^2\Biggr].
  \end{eqnarray}

 \subsection{Gaugino production}\label{gaugino_production}
  We consider the following seven processes in hadronic collisions:
  \begin{eqnarray}
   \label{g1} q\bar{q}  &\to& \ino\chi_a^+  \ino\chi_b^-   \\
   \label{g2} q\bar{q}  &\to& \ino\chi_i^0  \ino\chi_j^0   \\
   \label{g3} q\bar{q}' &\to& \ino\chi_a^\pm\ino\chi_i^0 \\   
   \label{g5} q\bar{q}  &\to& \ino\chi_i^0   \gluino         \\
   \label{g6} gq        &\to& \ino\chi_i^0   \sqk \,{\rm\ and\ } c.c. \\
   \label{g7} q\bar{q}' &\to& \ino\chi_a^\pm \gluino       \\
   \label{g8} gq        &\to& \ino\chi_a^\pm \sqk'\,{\rm\ and\ } c.c.
  \end{eqnarray}
  $c.c.$ refers to charge conjugation.
  The indices take the values $a,b=1,2$ and $i,j=1\ldots4$.
  Only processes (\ref{g1}) and (\ref{g2}) are relevant to leptonic
collisions, where the colour factors and electroweak coupling coefficient
need trivial modifications.
  Processes (\ref{g1}), (\ref{g2}), (\ref{g3}), (\ref{g5}) and (\ref{g7})
are expressed conveniently following the notation of Ref.~\cite{BRKP}:
  \begin{equation}\label{caba}A=\sum_{\alpha,\beta=L,R}\cab
   (\bar{v}_2\gamma^\mu{P}_\alpha u_1)
   (\bar{u}_3\gamma^\mu{P}_\beta v_4)
   =\sum_{\alpha,\beta=L,R}\cab Q_\alpha^\mu X_\beta^\mu,
  \end{equation}
  where, as before, ${P}_L=(1-\gamma_5)/2$ and ${P}_R=(1+\gamma_5)/2$.
  The spin-averaged amplitudes are given as:
  \begin{eqnarray}\label{cab}
   \overline{|A|^2}&=&[|C_{LL}|^2+|C_{RR}|^2]u_3u_4   
   +[|C_{LR}|^2+|C_{RL}|^2]t_3t_4+\nonumber\\
   &&+2\mathrm{Re}[C_{LL}^*C_{LR}+C_{RR}^*C_{RL}](m_3m_4s).
  \end{eqnarray}
  The masses appearing above as $(m_3m_4s)$ are the signed masses.

  The matrix elements squared for processes (\ref{g6}) and (\ref{g8}) have
the following form, where $\alpha$ labels the chirality of the final state
squark:
  \begin{equation}\label{classIII}
  \M_\alpha = \frac{(g_sg_\alpha)^2}{2N_C}
       \left[-\frac{t_3}{s}-\frac{2(m_4^2-m_3^2)t_3}{su_4}
            \left(1+\frac{m_3^2}{t_3}+\frac{m_4^2}{u_4}\right)\right].  
  \end{equation}
  When left--right squark mixing is considered, the above is modified as:
  \begin{eqnarray}
  \M_\alpha \propto g_\alpha^2 &\longrightarrow& \M_i \propto g_i^2,
  \rm{\ where} \nonumber\\  g_i^2 &=& g_L^2Q_{Li}^2+g_R^2Q_{Ri}^2.
  \end{eqnarray}

  Process (\ref{g1}), $ q\bar{q} \to \ino\chi_a^+\ino\chi_b^- $, is given
by eqn.~(\ref{cab}) with the prefactor $(e^4/N_C)$ and the following
coefficients \cab:
  \begin{eqnarray}
  \cab &=& -\delta_{ab}\frac{Q_q}{s}-\frac1{\sin2\theta_W}\cdot
           \frac{(\delta_{\alpha L}L_q+\delta_{\alpha R}R_q)
                 (-2O'^\beta_{ab})}\zz\nonumber\\
       &-& \frac{\delta_{\alpha L}}{2\sin^2\theta_W}
           \left[\frac{(1-\dab)\delta_{qu}U_{a1}U_{b1}}{t-M_{\sdn_L}^2}
           -\frac{\dab\delta_{qd}V_{a1}V_{b1}}{u-M_{\sup_L}^2}\right].
  \label{g1a}\end{eqnarray}
  Here, following the notation of Ref.~\cite{Haber:1985rc}, we have
defined:
  \begin{eqnarray}
      -2O'^L_{ab} &=& -V_{a2}V_{b2}+2\delta_{ab}\cos^2\theta_W;   \\
      -2O'^R_{ab} &=& -U_{a2}U_{b2}+2\delta_{ab}\cos^2\theta_W.
  \end{eqnarray}
   Process (\ref{g2}), $ q\bar{q} \to \ino{\chi}^0_i\ino{\chi}^0_j $, is
given by eqn.~(\ref{caba}) with the prefactors $(e^4/N_C)$ and
${1/(1+\delta_{ij})}$, and the following coefficients \cab:
  \begin{eqnarray}
  C_{LL} &=& \frac{L_q}{\sin2\theta_W}\cdot
             \frac{N_{i4}N_{j4}-N_{i3}N_{j3}}\zz+
             \frac{\ino{L}_{qi}\ino{L}_{qj}}{u-M_{\ino{q}_L}^2}; \\
  C_{LR} &=& \frac{L_q}{\sin2\theta_W}\cdot
             \frac{N_{i3}N_{j3}-N_{i4}N_{j4}}\zz-
             \frac{\ino{L}_{qi}\ino{L}_{qj}}{t-M_{\ino{q}_L}^2}; \\
  C_{RL} &=& \frac{R_q}{\sin2\theta_W}\cdot
             \frac{N_{i4}N_{j4}-N_{i3}N_{j3}}\zz-
             \frac{\ino{R}_{qi}\ino{R}_{qj}}{t-M_{\ino{q}_R}^2}; \\
  C_{RR} &=& \frac{R_q}{\sin2\theta_W}\cdot
             \frac{N_{i3}N_{j3}-N_{i4}N_{j4}}\zz+
             \frac{\ino{R}_{qi}\ino{R}_{qj}}{u-M_{\ino{q}_R}^2}.
  \label{g2a}\end{eqnarray}
 
   Process (\ref{g3}), $ u\bar{d} \to \ino{\chi}_a^+\ino{\chi}_i^0 $, is
given by eqn.~(\ref{caba}). The prefactors are given by $(e^4/N_C)$ and
$(\ckm/2\sin^2\theta_W)$ where $V_{\rm CKM}$ refers to the relevant term
of the Cabbibo--Kobayashi--Maskawa (CKM) matrix, and the coefficients
\cab\ are:
  \begin{eqnarray}
  C_{LL} &=& \frac1{\sin\theta_W}\cdot
             \frac{N_{i2}V_{a1}-N_{i4}V_{a2}/\sqrt2}\zw+
             \frac{V_{a1}\ino{L}_{ui}}{u-M_{\sup_L}^2}; \\
  C_{LR} &=& \frac1{\sin\theta_W}\cdot
             \frac{N_{i2}U_{a1}+N_{i3}U_{a2}/\sqrt2}\zw-
             \frac{U_{a1}\ino{L}_{di}}{t-M_{\sdn_L}^2}; \\
  C_{RL} &=& 0;  \\
  C_{RR} &=& 0. \end{eqnarray}

   Process (\ref{g5}), $ q\bar{q} \to \ino{\chi}_i^0\gluino $, is given by
eqn.~(\ref{caba}) with the prefactor $(e^2g_s^2C_F/N_C)$ and the following
coefficients \cab:
  \begin{eqnarray}
  C_{LL} &=&  \frac{\ino{L}_{qi}}{u-M_{\ino{q}_L}^2}; \\
  C_{LR} &=& -\frac{\ino{L}_{qi}}{t-M_{\ino{q}_L}^2}; \\
  C_{RL} &=&  \frac{\ino{R}_{qi}}{t-M_{\ino{q}_R}^2}; \\
  C_{RR} &=& -\frac{\ino{R}_{qi}}{u-M_{\ino{q}_R}^2}. \end{eqnarray}

   Process (\ref{g6}), $ gq \to \ino{\chi}^0_i\ino{q} $, is given by
eqn.~(\ref{classIII}), with $g_\alpha=e(\delta_{\alpha
L}\ino{L}_{qi}-\delta_{\alpha R} \ino{R}_{qi})$.
            
   Process (\ref{g7}), $ u\bar{d} \to \ino{\chi}_a^+\gluino $, is given by
eqn.~(\ref{caba}). The prefactors are $(e^2g_s^2C_F/N_C)$ and
$(\ckm/2\sin^2\theta_W)$, and the coefficients \cab\ are:
  \begin{eqnarray}
  C_{LL} &=&  \frac{V_{a1}}{u-M_{\ino{u}_L}^2}; \\
  C_{LR} &=& -\frac{U_{a1}}{t-M_{\ino{d}_L}^2}; \\
  C_{RL} &=&  0; \\
  C_{RR} &=&  0. \end{eqnarray}

   Process (\ref{g8}), $ gq \to \ino{\chi}^\pm_a\ino{q}'_L $, is given by
eqn.~(\ref{classIII}), with $g_L=e(\delta_{qu}U_{a1}+\delta_{qd}V_{a1})$.

   Processes (\ref{g3}), (\ref{g7}) and (\ref{g8}) involve
flavour-changing CKM matrix contributions. We assume that the CKM matrix
is universal and applicable also to sfermion interactions.

   Process (\ref{g1}) involves two flavour-changing vertex contributions
and, in principle, we should sum over the internal squark flavours and
external quark pairs. However, we believe that to a good approximation
this process can be treated correctly by substituting an identity matrix
for the CKM matrix and so this is done in \HW.

 \subsection{Neutral Higgs production}\label{neutral_Higgs}

  For hadronic collisions, the following five classes of production 
subprocesses are implemented:
  \begin{itemize}
  \item[\mbox{(a)}]  gluon-gluon fusion via loops 
  of heavy quarks $Q=b$ and $t$ and squarks $\ino{Q}=\ino{b}$ and 
  $\ino{t}$,
  \item[\mbox{(b)}]  $W^\pm W^\mp$, $Z^0Z^0$ fusion,
  \item[\mbox{(c)}]  associated production with $W^\pm$, $Z^0$ bosons,
  \item[\mbox{(d)}]  associated production with pairs of heavy quarks 
  $Q\bar Q$,
  \item[\mbox{(e)}]  associated production with pairs of heavy squarks
  $\ino{Q}\ino{Q}^*$.
  \end{itemize}

  The scalar Higgs bosons $h^0$ and $H^0$ can be produced by all five
subprocesses (a)--(e), whereas the pseudoscalar $A^0$ is only produced via
the first and the last two, as the $A^0VV$ vertex, with $V=W^\pm,Z^0$, is
absent at tree level. Higgs bosons are also produced in decays of heavier
sparticles.

  For leptonic collisions, the following are implemented:
  \begin{itemize}
  \item[\mbox{(f)}]  $W^\pm W^\mp$, $Z^0Z^0$ fusion,
  \item[\mbox{(g)}]  associated production with $Z^0$,
  \item[\mbox{(h)}]  $h^0A^0$ and $H^0A^0$ production.
  \end{itemize}
  Of these, (f) and (g) are expressed by formulae that are obtained by
changing the $Z^0$ boson couplings in the corresponding hadronic
subprocesses. In addition, for (g), the colour factor $1/N_C$ is replaced
by $1$.

  In the following, let us denote the neutral Higgs bosons collectively by
$\Phi$. We define the following notation for invariant mass squared. {\ims
ab } is the invariant mass squared of the particle pair $[ab]$, ${\ims
a_ib_i } ={\ims12 }=(p_a+p_b)^2=s$, $\ims a_fb_f =(p_a+p_b)^2$ and $\ims
a_ib_f =(p_a-p_b)^2$, where the subscripts $i$ and $f$ indicate particles
in the initial and final states, respectively. $\ims ab $ is negative only
in the last of these three cases.

  For the gluon-fusion subprocess class (a), the matrix elements are cast
in the form seen in Ref.~\cite{spira}:

  \begin{equation}\label{ME_ggh}
  \M({gg\to h^0,H^0}) = \frac{\alpha_{\rm{EM}}\as^2M_{h,H}^4 }{72\pi
\sin^2\theta_W (N_C^2-1)\mw^2} \ \left| \sum_{f} g_f^{h,H} A_f^{h,H}
(\tau_{f}) + \sum_{\ino{f}} g_{\ino{f}}^{h,H} A_{\ino{f}}^{h,H}
(\tau_{\ino{f}}) \right|^{2}, \end{equation}
\begin{equation}\label{ME_ggA} \M(gg\to A^0) =
\frac{\alpha_{\rm{EM}}\as^2M_A^4}{72\pi \sin^2\theta_W (N_C^2-1) \mw^2} \
\left| \sum_{f} g_f^A A_f^A (\tau_{f}) \right|^{2}, \end{equation}
 for scalar and pseudoscalar production, respectively. The couplings $g_f$
and $g_{\tilde f}$ are those appearing in Tabs.~4 and 5, respectively. We
have defined $\tau=4m^2/M^2$, where $M$ is the mass of the Higgs boson
concerned and $m$ is the mass of the particle inside the loop. The
functions $A$ are given by:
 \begin{eqnarray}\label{amp_h}
A_Q^{h,H}(\tau)&=&\frac{3}{2}\tau[1+(1-\tau)f(\tau)],\\
A_{\widetilde{Q}}^{h,H}(\tau)&=&-\frac{3}{4}[1-\tau f(\tau)],
 \end{eqnarray}
 for the scalars and
 \begin{equation}\label{amp_a} A_Q^{A}(\tau)=\tau f(\tau), \end{equation}
 for the pseudoscalar, with the function $f(\tau)$ given by
 \begin{eqnarray}\label{ftau}
f(\tau) & = & \left\{ \begin{array}{ll} \displaystyle \arcsin^2
\frac{1}{\sqrt{\tau}} & \tau \ge 1, \\ \displaystyle - \frac{1}{4} \left[
\log \frac{1+\sqrt{1-\tau}} {1-\sqrt{1-\tau}} - i\pi \right]^2
\hspace*{0.5cm} & \tau < 1. \end{array} \right. \end{eqnarray}

 \begin{center}
 \small
 \begin{tabular}{|lc|ccc|}
 \hline
\multicolumn{2}{|c|}{Higgs boson} & $g_{u,\nu}$ & $g_{d,\ell}$ &  
$g_{W^\pm,Z^0}$ \\ \hline \hline 
\SM~ & $\phi$ & 1 & 1 & 1 \\ \hline
\MSSM~ & $h^0$ & $\cos\alpha/\sin\beta$ & $-\sin\alpha/\cos\beta$ &
$\sin(\beta-\alpha)$ \\
& $H^0$ & $\sin\alpha/\sin\beta$ & $\cos\alpha/\cos\beta$ &
$\cos(\beta-\alpha)$ \\
& $A^0$ & $ 1/\tan\beta$ & $\tan\beta$ & 0 \\ \hline
 \end{tabular}
 \end{center}
 {{\bf Table 4:} Higgs coupling coefficients in the \MSSM\ to isospin
$+1/2$ and $-1/2$ fermions and to the gauge bosons $W^\pm,Z^0$, as first
appearing in eqns.~(\ref{ME_ggh}) and (\ref{ME_ggA}). }
   
 \begin{center}
 \small
 \begin{tabular}{|lc|c|}
 \hline
\multicolumn{2}{|c|}{Higgs boson} & $g_{\ino{f}_i}$ \\
\hline \hline
\SM~ & $\phi$ & 0 \\ \hline \rule[-10mm]{0cm}{23mm}
\MSSM~ & $h^0$ & $\begin{array}{c}{m^{-2}_{\ino{f}_i}}\Bigl[m_f^2g_f^h+
\mz^2\cos\theta_W\sin\theta_W(L_fQ_{Li}^2-R_fQ_{Ri}^2)\sin(\alpha+\beta)\\+
(\mu m_f(\delta_{fu}\frac{\sin\alpha}{\sin\beta}-
         \delta_{fd}\frac{\cos\alpha}{\cos\beta})+A_fm_fg^h_f)Q_{Li}Q_{Ri}
\Bigr]\end{array}$ \\ \rule[-10mm]{0cm}{20mm}
      & $H^0$ & $\begin{array}{c}{m^{-2}_{\ino{f}_i}}\Bigl[m_f^2g_f^H+ 
\mz^2\cos\theta_W\sin\theta_W(R_fQ_{Ri}^2-L_fQ_{Li}^2)\cos(\alpha+\beta)\\+
(\mu m_f(-\delta_{fu}\frac{\cos\alpha}{\sin\beta}-
         \delta_{fd}\frac{\sin\alpha}{\cos\beta})+A_fm_fg^H_f)Q_{Li}Q_{Ri}
\Bigr]\end{array}$ \\\rule[-5mm]{0cm}{10mm}
& $A^0$ & 0 \\ \hline
 \end{tabular}
 \end{center}
 {{\bf Table 5:} \MSSM\ Higgs coupling coefficients to sfermions ${\tilde
f}={\tilde q},{\tilde \ell}$, as first appearing in eqn.~(\ref{ME_ggh}).
For the leptonic case, the mixing matrices $Q_{ij}$ should be replaced by
the corresponding slepton mixing matrices $L_{ij}$. }

For process (b), the matrix element for the $W^\pm W^\mp$ fusion process
 is given by:
\begin{eqnarray}\label{qqWW}
{\M}(q_1q'_2\to q''_3q'''_4\Phi_5) & = &  ({g_{W^\pm}^{\Phi}})^2~
\left(\frac{e^2}{2\sin^2\theta_W}\right)^3(2\mw^2)\nonumber\\ &\times&
   \frac{\ckmc{qq''}
   \ckmc{q'q'''}\ims12 \ims34 }{(\ims13 -\mw^2)(\ims24 -\mw^2)},
\end{eqnarray}
whereas for $Z^0Z^0$ fusion we have:
\begin{eqnarray}\label{qqZZ}
{\M}(q_1q'_2\to q_3q'_4\Phi_5) & = &
({g_{Z^0}^{\Phi}})^2 
\left(\frac{e^2}{4\sin^2\theta_W\cos^2\theta_W}\right)^3(4\mz^2)
\nonumber\\ &\times&
\frac{(L^2_{q }L^2_{q'}+R^2_{ q}R^2_{q'})\ims12 \ims 34 + 
                (L^2_{q }R^2_{q'}+R^2_{ q}L^2_{q'})\ims14 \ims 23 }
               {(\ims13 -\mz^2)(\ims24 -\mz^2)}.\quad
\end{eqnarray}
 The coupling coefficients $g_V^\Phi$ are as given in Tab.~4. The formulae
are identical to those for the case of leptonic collisions, with the
replacement of the CKM matrix elements by unity for the $W^\pm W^\mp$
fusion subprocess and the substitution of the couplings $L,R$ with the
appropriate leptonic couplings $L_e,R_e$ for the $Z^0Z^0$ fusion
subprocess.

 For process (c), the matrix elements are given by:
 \begin{equation}\label{WH}
{\M}(u_1\bar{d}_2\to W^+_3\Phi_4)  =  ({g_{W^\pm}^{\Phi}})^2
\frac{1}{N_C}\left(\frac{e^2}{2\sin^2\theta_W}\right)^2
                       \left(\frac{1}{2}\right)\ckmc{ud}
          \frac{sp_T^2+2\mw^2s}{(s-\mw^2)^2+\mw^2\Gambr_W^2},
 \end{equation}
 \begin{equation}\label{ZH}
{\M}(q_1\bar{q}_2\to Z_3\Phi_4)  =  ({g_{Z^0}^{\Phi}})^2
\frac{1}{N_C}\left(\frac{e^2}{4\sin^2\theta_W\cos^2\theta_W}\right)^2
          (L_q^2+R_q^2)
          \frac{sp_T^2+2\mz^2s}{(s-\mz^2)^2+\mz^2\Gambr_Z^2}.
 \end{equation}
 The coupling coefficients $g_V^\Phi$ are again as given in Tab.~4. The
above formula, for the $Z^0$ mediated subprocess, is immediately
applicable to the leptonic case with the substitution of the colour factor
$1$ to replace $1/N_C$. The couplings $L_q,R_q$ are replaced by $L_e,R_e$
in this case.

  For process (d), the matrix element for the $q\bar q$ initiated
subprocess is given by:
 \begin{eqnarray}{\M}(q_1\bar{q}_2\to Q_3\bar{Q}_4\Phi_5) &=&
   \frac{e^2g_s^4(g_Q^\Phi)^2}{2\sin^2\theta_W\mw^2s}
   \left(\frac{C_F}{N_C}\right)\prop\nonumber\\
   &\times&\left[
    \frac{1}{2\prop_5}-\frac{M^2s\prop}{2}+
    M^2(\prop_3{p^2_T}_3+\prop_4{p^2_T}_4)-\prop_5(s+M^2){p^2_T}_5
   \right].\nonumber\\ \end{eqnarray}
  Here $M^2=M_\Phi^2$ for $A^0$ and $M=M_\Phi^2-4m_Q^2$ for $h^0, H^0$.
The $p_T$'s are the transverse momenta of the final state particles. These
can be expressed as $p_{T_i}^2=4p_1\cdot p_i\;  p_2\cdot p_i/s-m_i^2$
(where $i=3,4,5$) as the incoming particles are considered massless. The
propagator factors are defined by:
 \begin{eqnarray}
 \prop_3 &=& 1/(\ims35 -m_Q^2), \\
 \prop_4 &=& 1/(\ims45 -m_Q^2), \\
 \prop   &=& \prop_3+\prop_4,   \\
 \prop_5 &=& \prop_3\prop_4/\prop.
 \end{eqnarray}

  For the case $Q=b$, the $gg$ initiated subprocess is evaluated with
$2\to1$, $2\to2$ and $2\to3$ matrix elements. In order to avoid double
counting, it is recommended that the user chooses only one of these,
rather than sum over them, according to their need. For the $2\to1$
subprocess the matrix element is given by:
 \begin{equation}\label{bbH}
  {\M}(b_1\bar b_2\to\Phi_5) = \frac{e^2M_H^2m_b^2(g_b^\Phi)^2}
                                    {8\sin^2\theta_W\mw^2N_C}.
 \end{equation}

  The $2\to2$ matrix element is given by:
  \begin{eqnarray}\label{gbbH}
  \M(g_1b_2\to b_3\Phi_4) &=& \left(\frac{g_s^2}{2N_C}\right)
   \left(\frac{e^2}{2\sin\theta^2_W}\right)
   \left(\frac{m_b^2(g_b^\Phi)^2}{2\mw^2}\right)
    \nonumber \\&\times&\left(\frac{-u_4^2}{st_3}\right)
     \left[1+2\frac{m_4^2-m_3^2}{u_4}
      \left(1+\frac{m_3^2}{t_3}+\frac{m_4^2}{u_4}\right)\right].
  \end{eqnarray}
  The inconsistency in treating the initial state bottom quark as being
massless and the final state bottom quark as being massive is unavoidable.  
The subprocess implementation being intended for high $p_T$ single $b$-jet
events only, the ambiguity in the treatment of the bottom quark mass in
the hard scattering matrix element is negligible.


  We evaluated the $2\to3$ matrix element for $gg\to Q\bar{Q}\Phi$ using
helicity amplitudes as follows. We first wrote down the amplitude
functions using the corresponding Feynman rules. We then adopted the
convention of Ref.~\cite{Kleiss:1985yh} for the heavy quark helicity
eigenstates, and adopted the circular polarization for the gluons. This
allowed some simplification in the computation which, together with the
factorization of amplitudes, made the code very much faster than any other
code which is publically available.

  For process (e), the matrix element for the $q\bar q$ initated
subprocess is given by:
  \begin{eqnarray}
  \M(q_1\bar q_2\to\ino{Q}_3\ino{Q}^*_4\Phi_5)&=&\left(
  \frac{g_S^4e^2{g^\Phi_{\ino{Q}}}^2M_{\ino{Q}}^2}{\sin^2\theta_W\mw^2
  s}\right)\left(\frac{C_F}{N_C}\right)\nonumber\\\label{qsqH}
  &\times&\left[|\prop_3|^2{p^2_T}_3+|\prop_4|^2{p^2_T}_4-
  2\mbox{Re}\left(\prop_3^*\prop_4\right){p_T}_3\cdot{p_T}_4\right].
  \end{eqnarray}
  The propagator factors are given by:
  \begin{eqnarray}
  \prop_3&=&1/\left((p_4+p_5)^2-m_3^2+im_3\Gambr_3\right),\\
  \prop_4&=&1/\left((p_3+p_5)^2-m_4^2+im_4\Gambr_4\right).
  \end{eqnarray}
  The dot product of two transverse momenta is two dimensional, \ie\
${p_T}_3\cdot{p_T}_4={p_x}_3{p_x}_4+{p_y}_3{p_y}_4$.

  The $gg$ initiated subprocess is again calculated using helicity
amplitudes in an optimized basis. This time the two orthogonal
polarisation vectors for the initial state gluons are chosen to be in the
$x$ and $y$ directions, yielding a relatively compact expression:
  \begin{eqnarray}
  \M(g_1g_2\to\ino{Q}_3\ino{Q}^*_4\Phi_5)&=&\left(
  \frac{g_S^4e^2{g^\Phi_{\ino{Q}}}^2M_{\ino{Q}}^2N_C}{4\sin^2\theta_W\mw^2
  (N_C^2-1)}\right)\nonumber\\&\times&\sum_{i=x,y}\sum_{j=x,y}
  \left[|{\mathcal A}_t(i,j)|^2+|{\mathcal A}_u(i,j)|^2-
  \frac1{N_C^2}|{\mathcal A}_t(i,j)+{\mathcal A}_u(i,j)|^2\right].
  \nonumber\\\label{gsqH}
  \end{eqnarray}
  The colour flow components are distributed using the method of
Ref.~\cite{Odagiri:1998ep}, namely, the $(1/N_C^2)$ term is distributed
amongst the $t$-channel and $u$-channel colour flow components by their
ratio. The two amplitude functions $\mathcal A_t$ and $\mathcal A_u$ are
given by:
  \begin{eqnarray}
  \mathcal{A}_t(i,j)&=&2\left({p_i}_3{p_j}_4\prop_{24}\prop_{13}+
  {p_i}_3{p_j}_3\prop_{13}\prop_{45}+{p_i}_4{p_j}_4\prop_{24}\prop_{35}
  \right)\nonumber\\&-&\frac{\delta_{ij}}{\sqrt{s}}
  \left({p_z}_3\prop_{45}-{p_z}_4\prop_{35}\right)-
  \frac{\delta_{ij}}2\left(\prop_{35}+\prop_{45}\right),\\
  \mathcal{A}_u(i,j)&=&2\left({p_i}_4{p_j}_3\prop_{14}\prop_{23}+
  {p_i}_3{p_j}_3\prop_{23}\prop_{45}+{p_i}_4{p_j}_4\prop_{14}\prop_{35}
  \right)\nonumber\\&+&\frac{\delta_{ij}}{\sqrt{s}}
  \left({p_z}_3\prop_{45}-{p_z}_4\prop_{35}\right)-
  \frac{\delta_{ij}}2\left(\prop_{35}+\prop_{45}\right).
  \end{eqnarray}
  The propagator factors are given by:
  \begin{eqnarray}
  \prop_{13}&=&1/\left((p_1+p_3)^2-m_3^2\right),\\
  \prop_{23}&=&1/\left((p_2+p_3)^2-m_3^2\right),\\
  \prop_{14}&=&1/\left((p_1+p_4)^2-m_4^2\right),\\
  \prop_{24}&=&1/\left((p_2+p_4)^2-m_4^2\right),\\
  \prop_{35}&=&1/\left((p_3+p_5)^2-m_4^2+im_4\Gambr_4\right),\\
  \prop_{45}&=&1/\left((p_4+p_5)^2-m_3^2+im_3\Gambr_3\right).
  \label{lastgsqH}
  \end{eqnarray}

  For process (h), the matrix element is given by:
  \begin{equation}
  \M(e_1^- e_2^+\to h^0_3A^0_4,H^0_3A^0_4)=
  e^4sp_T^2\cdot\frac{L_\nu^2\left(L_e^2+R_e^2\right)
  \left(\cos^2(\beta-\alpha),\sin^2(\beta-\alpha)\right)}
  {(s-\mz^2)^2+\mz^2\Gambr_Z^2}.\label{qqha}
  \end{equation}
  We have used the same notation as in eqn.~(\ref{ee_spartons}). The
coupling factor $\cos^2(\beta-\alpha)$ is for $h^0A^0$ production whereas
$\sin^2(\beta-\alpha)$ is for $H^0A^0$ production.

 \subsection{Charged Higgs production}

  For hadronic collisions, the following production modes are implemented:
  \begin{eqnarray}
  gb      &\to& t H^- + c.c.;                \label{gb_th}\\
  gg      &\to& t \bar b H^- + c.c.;         \label{gg_tbh}\\
  q\bar q &\to& t \bar b H^- + c.c.;         \label{qq_tbh}\\
  gg      &\to& \ino{t}\ino{b}^* H^- + c.c.; \label{gg_sqh}\\
  q\bar q &\to& \ino{t}\ino{b}^* H^- + c.c.; \label{qq_sqh}\\
  q\bar q &\to& H^+H^-;                      \label{qq_hh}\\
  b\bar b &\to& W^\pm H^\mp;                 \label{bb_wh}\\
  qb      &\to& q'bH^+ + c.c.                \label{qb_qbh}
  \end{eqnarray}
  Charged Higgs bosons are also produced through the decay of heavier
objects including gauginos and third generation sfermions. Perhaps most
notably, an important potential production mode for charged Higgs bosons
is through the decay of top quarks and, as with other decay processes,
this is achieved by specifying this decay mode in the input file. The
decay is carried out using the three-body matrix element. This matrix
element is not included in the spin correlation algorithm
\cite{Richardson:2001df} which is the default treatment from version 6.4.
If this decay is required, the matrix element code {\tt NME=200} should be
used and spin correlations and supersymmetric three body decays should be
switched off, \ie\ {\tt SYSPIN=.FALSE.} and {\tt THREEB=.FALSE.}.

  When simulating processes (\ref{gg_tbh}), (\ref{qq_tbh}) and
(\ref{qb_qbh}) one should bear in mind that there is potentially a large
contribution from top quark decay when this decay mode is available. In
this case the user should choose to simulate either the top quark
production or the relevant three-body process in order to avoid double
counting. In the limit in which the bottom quark is generated by the
collinear splitting $g\to b\bar b$, process (\ref{gg_tbh}) is described by
the two body process (\ref{gb_th}). This contribution becomes significant
if the mass of the charged Higgs boson is comparable or greater than the
top quark mass. Again, the user should choose to simulate either of the
two processes in order to avoid double counting.

  For leptonic collisions, in addition to the top quark decay, charged
Higgs bosons can be produced in a process analogous to process
(\ref{qq_hh}):
  \begin{equation}
  e^-e^+ \to H^+H^-. \label{ee_hh}
  \end{equation}
  The matrix element is obtained immediately from the hadronic case by
changing the appropriate photon and $Z^0$ couplings and removing the
colour factor $1/N_C$.

  Process (\ref{gb_th}) is given by:
  \begin{eqnarray}\label{gb0}
  \M(g_1b_2\to t_3H^-_4) &=& \left(\frac{g_s^2}{2N_C}\right)
   \left(\frac{e^2}{2\sin\theta^2_W}\right)\ckmc{tb}
   \left(\frac{m_b^2\tan^2\beta+m_t^2\cot^2\beta}{2\mw^2}\right)
    \nonumber \\&\times&\left(\frac{-u_4^2}{st_3}\right)
     \left[1+2\frac{m_4^2-m_3^2}{u_4}
      \left(1+\frac{m_3^2}{t_3}+\frac{m_4^2}{u_4}\right)\right].
  \end{eqnarray}

  Process (\ref{gg_tbh}) is evaluated using helicity amplitudes in an
optimized basis, similarly to the evaluation of the neutral Higgs boson
production process $gg\to Q\bar Q\Phi$ in Sect.~\ref{neutral_Higgs}.

  Process (\ref{qq_tbh}) is evaluated using a more general formula for
$q_1\bar q_2\to Q_3\bar Q_4\Phi$ which is given by:
  \begin{eqnarray}
  \M(q_1\bar q_2\!\to\! Q_3\bar Q_4\Phi)
   &=& \left(\frac{g_s^4C_F}{2N_C}\right)
   \frac{2\phi_H^2}s
   \Biggl[\left(p_{T_3}^2+p_{T_4}^2-p_{T_5}^2\right)+
          4p_0\cdot p_3\ p_0\cdot p_4 |\prop|^2 \nonumber\\
   &+& 2\left(p_3\cdot p_4\!-\!\lambda m_3m_4\right)\left[-s|\prop|^2+
       2\mathrm{Re}\left(\prop\prop_3^*p_{T_4}^2+\prop\prop_4^*p_{T_3}^2
       -\prop_3\prop_4^*p_{T_5}^2\right)\right]\nonumber\\
   &-& \!4\mathrm{Re}\!\!
       \left[\prop\prop_3^*p_{T_4}^2+\prop\prop_4^*p_{T_3}^2
       +\prop\left(\prop_3^*p_0\cdot p_4+\prop_4^*p_0\cdot p_3\right)
       \frac{p_{T_3}^2+p_{T_4}^2-p_{T_5}^2}2\right]\nonumber\\
   &-& 4\mathrm{Im}\left[\prop\prop_3^*p_{z_4}-\prop\prop_4^*p_{z_3}\right]
       \lambda'\sqrt{s}\left(p_{x_4}p_{y_3}-p_{x_3}p_{y_4}\right)\Biggr],
  \end{eqnarray}
  with $p_0=p_1+p_2$. Denoting the $Q_3\bar Q_4\Phi$ vertex as
$\left(g_\mathrm{S}+g_\mathrm{P}\gamma_5\right)$, the various
coupling coefficients appearing above are defined by:
  \begin{eqnarray}
   \phi_H^2 &=& g_\mathrm{S}^2+g_\mathrm{P}^2,\\
   \lambda  &=& \frac{g_\mathrm{S}^2-g_\mathrm{P}^2}{\phi_H^2},\\
   \lambda' &=& \frac{2g_\mathrm{S}g_\mathrm{P}}{\phi_H^2}.
  \end{eqnarray}

  Process (\ref{gg_sqh}) is evaluated using helicity amplitudes. The
expressions are the same as eqns.~(\ref{gsqH})--(\ref{lastgsqH}) but the
coupling coefficient needs to be changed. Let us consider the vertex
involving an incoming $H^+$ and outgoing $\ino{t}_i\ino{b}_j^*$. We have
the following replacement:
  \begin{eqnarray}
  \frac{g_{\ino{Q}}M_{\ino{Q}}^2}{\mw^2}\to
  \frac1{\sqrt{2}}&\Biggl[&
  \left(\sin2\beta-\frac{m_b^2\tan\beta+m_t^2\cot\beta}{\mw^2}\right)
  Q^t_{Li}Q^b_{Lj}-
  \frac{m_tm_b}{\mw^2}\left(\tan\beta+\cot\beta\right)Q^t_{Ri}Q^b_{Rj}
  \nonumber\\&-&
  \frac{m_b}{\mw^2}\left(-\mu+A_b\tan\beta\right)Q^t_{Li}Q^b_{Rj}
  -
  \frac{m_t}{\mw^2}\left(-\mu+A_t\cot\beta\right)Q^t_{Ri}Q^b_{Lj}
  \Biggr].
  \end{eqnarray}

  Process (\ref{qq_hh}) is given by:
  \begin{eqnarray}
  \M &=&
  \frac{e^4sp_T^2}{N_C}\nonumber\\&\times&\left[
  \left|\frac{Q_qQ_{H^+}}s+\frac{L_qL_{H^+}}{\zz}\right|^2+
  \left|\frac{Q_qQ_{H^+}}s+\frac{R_qL_{H^+}}{\zz}\right|^2
  \right].\label{qqhphm}
  \end{eqnarray}
  We have used the same notation as in eqn.~(\ref{ee_spartons}).
$Q_{H^+}=+1$, and $L_{H^+}$ is given by eqn.~(\ref{def_Lf}) with
$I_f^3=+1/2$. For the leptonic case the subscripts $q$ should be replaced 
with $e$, and the colour factor $1/N_C$ should be replaced by $1$.

  The matrix element for process (\ref{bb_wh}) is as given in
\cite{BarrientosBendezu:1998gd}.


  Process (\ref{qb_qbh}) is given in \cite{Moretti:1996ra}. We use a more
compact expression which sets the kinematic bottom quark mass to zero. For
the process $b_1U_2\to b_3D_4H^+_5$ we have:
 \begin{eqnarray}
 \M(b_1U_2\to b_3D_4H^+_5)
    &=&\frac{1}{2}\left(\frac{e^2}{2x_W}\right)^3\ckmc{UD}
    \frac{1}{(\ims24 -M_W^2)^2}\times\nonumber\\&&\times
    \left[\ckmc{tb}\overline{|A_t|^2}+\overline{|A_\Phi|^2}+
    {\rm Re}(V_{\rm CKM}[tb]\cdot\overline{2A_\Phi^*A_t})\right];  \\
 \overline{|A_t|^2}&=&2\ims12 |\prop_t|^2
           \left[\left(\frac{m_bt_\beta}{m_W}\right)^2\frac{[3545]}{2}  
      +\left(\frac{m_tt^{-1}_\beta}{m_W}\right)^2m_t^2\ims34 \right]; \\
 \overline{|A_\Phi|^2}&=&(\prop_{h,H}^2+\prop_A^2)(-\ims13 )
           \frac{[2545]}{2};\\
 {\rm Re}(\overline{2A_\Phi^*A_t})&=&\frac{m_bt_\beta}{M_W}
  (\prop_{h,H}+\prop_A)
  \Bigl[2{\rm Re}(\prop_t)(\ims45 -M^2_{H^\pm})(\ims12 )(-\ims13 )
   -\nonumber\\&&-{\rm Re}(\prop_t)\ims45 [1243]
    + {\rm Im}(\prop_t)\ims45 \cdot 4\det(1234)\Bigr] .
 \end{eqnarray}
   Here we have defined $[abcd]=4(p_a\cdot p_b\;p_c\cdot p_d + p_a\cdot  
p_d\;p_b\cdot p_c - p_a\cdot p_c\;p_b\cdot p_d)$ and $\det(abcd)=
\det(p_a^\mu,p_b^\nu,p_c^\rho,p_d^\sigma)=\sum_{\mu\nu\rho\sigma} 
\epsilon_{\mu\nu\rho\sigma}p_a^\mu p_b^\nu p_c^\rho p_d^\sigma$ where 
$\epsilon_{0123}=+1$.

The propagators are defined by:
 \begin{eqnarray*}
 \prop_t    &=& \frac{1}{\ims35 -m_t^2+im_t\Gambr_t}
  =\frac{\ims35 -m_t^2-im_t\Gambr_t}{(\ims35 -m_t^2)^2+m_t^2\Gambr_t^2};\\
 \prop_{h,H}&=&
   \frac{m_b\sin\alpha\cos(\beta-\alpha)/M_W\cos\beta}{\ims13 -M_h^2} +
   \frac{m_b\cos\alpha\sin(\beta-\alpha)/M_W\cos\beta}{\ims13 -M_H^2}; \\
 \prop_A    &=& \frac{m_b\tan\beta/M_W}{\ims13 -M_A^2}.
 \end{eqnarray*}
   Processes with antiquarks in the initial state can be obtained by
exchanging the subscripts $(1\leftrightarrow3)$ for the subprocess
$\bar{b}U\to\bar{b}DH^+$, $(2\leftrightarrow4)$ for the subprocess
$b\bar{D}\to b\bar{U}H^+$, and both for the subprocess
$\bar{b}\bar{D}\to\bar{b}\bar{U}H^+$.
  Note that the definitions of the invariant
masses squared are such that these imply momentum substitutions of the
form $(p_1\leftrightarrow -p_3)$.

 \subsection{R-parity violating processes}


  The cross sections for the implemented R-parity violating hadron--hadron
processes are presented in Ref.~\cite{Dreiner:1999qz}. In hadron--hadron
collisions only those $2\to2$ processes that can occur via a resonant
$s$-channel sparticle exchange are included, \ie\ processes that can only
occur via a $t$-channel diagram or for which the $s$-channel resonance is
not kinematically accessible are neglected.

  In $e^+e^-$ collisions we include all the possible $2\to2$ single
sparticle production mechanisms, even if there is no $s$-channel
resonance, and a limited set of $2\to2$ processes where the final state
contains two \SM\ particles. As with the hadron--hadron processes, there
are four types of process: single gaugino production; single slepton
production in association with a gauge boson; single slepton production in
association with a Higgs boson; \SM\ particle production.

%
%
\subsubsection{Single gaugino production}

  In $e^+e^-$ collisions, there are two possible production mechanisms for
a gaugino and a lepton, either $e^+e^-\to\cht^0\nu$ or
$e^+e^-\to\cht^+\ell^-$.
  The matrix elements for these processes have been calculated previously
in Refs.~\cite{Dreiner:1995ij,Giudice:1996dm,Chemtob:1999uq}.
  We have recalculated these matrix elements and the results are in
agreement with \cite{Chemtob:1999uq}. We also agree with
Refs.~\cite{Dreiner:1995ij,Giudice:1996dm}, which do not include sfermion
mixing, if we set our mixing to zero.
  All matrix elements are averaged over the spins of the incoming leptons
as before.
  \begin{eqnarray}
\lefteqn{\displaystyle{
\overline{|\mathcal{M}|^2}(e_i^-e_i^+\rightarrow\nu_j\cht^0_l)=}} & \nonumber\\
&&\displaystyle{\frac{\lam^2_{iji}}{2}\left[\rule{0mm}{6.3mm} 
	|a(\nut_j^*)|^2sR(\nut_j)\left(s-M^2_{\cht^0_l}\right)
+\frac1{(u-M^2_{\ell_{iL}})^2}
   	\left[|a(\elt^*_{i1})|^2+|b(\elt^*_{i1})|^2\right]
	u\left(u-M^2_{\cht^0_l}\right)\right.}\nonumber\\
&&\displaystyle{+\frac1{(t-M^2_{\ell_{iR}})^2}
   	\left[|a(\elt_{i2})|^2+|b(\elt_{i2})|^2\right]
	t\left(t-M^2_{\cht^0_l}\right)}\nonumber\\
&&\displaystyle{+\frac2{(u-M^2_{\ell_{iL}})}a(\nut^*_j)a(\elt^*_{i1})
	R(\nut_j)\left(s-m^2_{\nut_j}\right)su
+\frac{2}{(u-M^2_{\ell_{iL}})(t-M^2_{\ell_{iR}})}a(\elt^*_{i1})a(\elt_{i2})ut}
\nonumber\\
&&\displaystyle{\left.+\frac2{(t-M^2_{\ell_{iR}})}a(\nut^*_j)a(\elt_{i2})
	R(\nut_j)\left(s-m^2_{\nut_j}\right)st
\rule{0mm}{6.3mm}\right],}\\
\lefteqn{\displaystyle{
\M(e_i^-e_i^+\rightarrow\ell^+_j\cht^-_l)=}} & \nonumber\\
&&\displaystyle{\frac{g^2\lam^2_{iji}}{4}\left[\rule{0mm}{7mm}
   R(\nut_j)s\left[\left(|a(\nut_j^*)|^2+|b(\nut_j^*)|^2\right)
   \left(s-m^2_{\ell_j}-M^2_{\cht^-_l}\right)
   -4a(\nut_j^*)b(\nut_j^*)m_{\ell_j}M_{\cht^-_l}\right]
\right.}\nonumber\\
&&\displaystyle{+\frac1{\left(u-M^2_{\nut_i}\right)^2}\left(M^2_{\cht^-_l}-u\right)
 	\left(m^2_{\ell_j}-u\right)\left[|a(\nut^*_i)|^2+|b(\nut^*_i)|^2\right]}
\nonumber\\
&&\displaystyle{\left.-\frac{2s}{\left(u-M^2_{\nut_i}\right)}
	\left(s-M^2_{\nut_j}\right)R(\nut_j)a(\nut^*_i)
	\left[a(\nut^*_j)u+m_{\ell_j}M_{\cht^-_l}b(\nut^*_j)
	\right]
\rule{0mm}{7mm}\right].}
\end{eqnarray}
  The couplings $a,b$ of the sleptons to the gauginos are given in
Ref.~\cite{Dreiner:1999qz}, and \begin{equation}
  R(\tilde{a}) = \frac1{\left(s-M^2_{\tilde{a}}\right)^2
			+\Gambr^2_{\tilde{a}}M^2_{\tilde{a}}}.
\end{equation}

%
%
\subsubsection{Slepton production in association with a gauge boson}

  There are a number of production processes for a slepton in
  association with a gauge boson. We have confirmed the cross section for
  these processes, which were first calculated in
Ref.~\cite{Chemtob:1999uq}.
  The sneutrino resonance is not usually
  accessible for these processes as the charged slepton--sneutrino mass
  difference is less than the $W^\pm$ mass. However, the resonance can be
accessible for the third generation due to left--right mixing of the
staus.

  The matrix elements for these processes are given below, including
left--right sfermion mixing:
\begin{eqnarray}
\lefteqn{\displaystyle{
\overline{|\mathcal{M}|^2}(e_i^-e_i^+\rightarrow\gamma\nut_j)=}}&
\nonumber\\
&&\displaystyle{\frac{e^2\lam_{iji}^2}{2}\left[\rule{0mm}{6mm}\frac{u}{t}+\frac{t}{u}+
   \frac2{ut}\left(M^2_{\nut_j}-u\right)\left(M^2_{\nut_j}-t\right)\right],}\\
\lefteqn{\displaystyle{
\overline{|\mathcal{M}|^2}(e_i^-e_i^+\rightarrow W^+\elt_{j\al})=}}&
\nonumber\\
&&\displaystyle{\frac{g^2\lam_{iji}^2|L^{j}_{1\al}|^2}{2\mw^2}\left[\rule{0mm}{6mm}
s^2p^2_{cm}R(\nut_j)+\frac1{4u^2}
    \left[2\mw^2\left(ut-\mw^2M^2_{\elt_{j\al}}\right)+u^2s\right]\right.}\nonumber\\
 &&\displaystyle{\left.  -\frac{s}{2u}R(\nut_j)\left(s-M^2_{\nut_j}\right)
   \left[\mw^2\left(2M^2_{\elt_{j\al}}-u\right)+u\left(s-M^2_{\elt_{j\al}}\right)
  \right]
\rule{0mm}{6mm}\right],}\\
\lefteqn{\displaystyle{
\overline{|\mathcal{M}|^2}(e_i^-e_i^+\rightarrow Z^0\nut_j)=}}&
 \nonumber\\
&&\displaystyle{\frac{g^2\lam_{iji}^2}{\cos^2\theta_W\mz^2}\left[\rule{0mm}{8mm}
    |Z^{11}_{\nu_j}|^2s^2p^2_{\rm{cm}}R(\nut_j)
    +\frac{Z^2_{e_R}}{t^2}\left[2\mz^2\left(ut-M^2_{\nut_j}\mz^2\right)
		+st^2\right]\right.}\nonumber\\
&&\displaystyle{+\frac{Z^2_{e_L}}{u^2}\left[2\mz^2\left(ut-M^2_{\nut_j}\mz^2\right)
	+su^2\right] +\frac{2Z_{e_L}Z_{e_R}}{ut}\left[2\mz^2\left(M^2_{\nut_j}-t\right)
  	\left(M^2_{\nut_j}-u\right)-sut\right]}\nonumber\\
&&\displaystyle{-\frac{Z^{11}_{\nut_j}Z_{e_R}}{t}sR(\nut_j)\left(s-M^2_{\nut_j}\right)
 	\left[\mz^2\left(2M^2_{\nut_j}-t\right)
	+t\left(s-M^2_{\nut_j}\right)\right]} \nonumber\\
&&\displaystyle{\left.  +\frac{Z^{11}_{\nut_j}Z_{e_L}}{u}sR(\nut_j)
	\left(s-M^2_{\nut_j}\right)\left[\mz^2\left(2M^2_{\nut_j}-u\right)
	+u\left(s-M^2_{\nut_j}\right)\right]\rule{0mm}{8mm}\right],}
\end{eqnarray}
  where $p_{\rm{cm}}$ is the momentum of the final-state particles
  in the centre--of--mass frame.
  As we neglect the electron mass,
  the cross section for $e^-_ie^+_i\rightarrow\gamma\nut_j$ is divergent
as $u,t\rightarrow0$, \ie\ in the limit that the photon is collinear with
the incoming lepton beams, and therefore a cut on the transverse momenta
of the outgoing particles is used to regularize the divergence for this
process.
 The couplings of the $Z^0$ to the leptons and sleptons are given in 
 Ref.~\cite{Dreiner:1999qz}.
%
%
\subsubsection{Slepton production in association with a Higgs boson}

  There are a number of processes in which a Higgs boson can be produced in
  association with a slepton. The cross sections for these processes tend to be
  small due to the small lepton masses and the inaccessibility of the 
  sneutrino resonance as the  charged slepton--sneutrino mass difference 
  is smaller than the charged Higgs mass. 
  
  The matrix elements are given below
\begin{eqnarray}
{\displaystyle\lefteqn{\M(e_i^-e_i^+\rightarrow \elt^*_{j\al} H^-) =}}\nonumber\\
 &{\displaystyle \frac{g^2 {\lam}_{iji}^2}{4}\left[
	 |H^c_{\nut\elt_{j\al}}|^2 s  R(\nut_j)
	+\frac{4|L^{2j-1}_{1\al}|^2|E^c_j|^2}{u^2}
	 \left(ut-M^2_{ \elt_{j\be}}M^2_{H^-}\right)\rule{0mm}{0.8cm}\right]\!,} \\
 & \nonumber \\
{\displaystyle \lefteqn{\M(e_i^-e_i^+\rightarrow \nut^*_j H_l^0) =}}
\nonumber\\
 & {\displaystyle \frac{g^2 {\lam}_{iji}^2}{4}\left[\rule{0mm}{0.8cm}
	|H^l_{\nut_{j}\nut_{j}}|^2s R(\nut_{j})
+\frac{|E^l_i|^2}{u^2}
		\left(ut-M^2_{ \nut_j}M^2_{H_l^0}\right)
 	  +\frac{|E^l_i|^2}{t^2}
		\left(ut-M^2_{\nut_j}M^2_{H_l^0}\right)\rule{0mm}{0.8cm} \right]\!.}
\end{eqnarray}
  The couplings $H$ of the Higgs bosons to the sleptons are given in 
Ref.~\cite{Dreiner:1999qz}
  and the couplings $E^l_i$ of the leptons to the Higgs bosons can be obtained from
  the couplings given in Ref.~\cite{Dreiner:1999qz} for the down-type 
quarks by replacing 
  the quark mass with the lepton mass.

%
%
\subsubsection{\SM\ particle production by sparticle exchange}

  As with the hadron-hadron processes described in
Ref.~\cite{Dreiner:1999qz} we have only included $2\to2$ processes where
two \SM\ particles can be produced by a resonant $s$-channel exchange.
Where this is possible we have included any $t$-channel sparticle
exchanges, and the relevant interference terms. We have also included the
relevant \SM\ diagrams and the interference between the \SM\ and the
R-parity violating diagrams. We have generalized the results of
Ref.~\cite{Kalinowski:1997bc}, giving
 \begin{eqnarray} \displaystyle{
\lefteqn{\overline{|\mathcal{M}|^2}(e_i^-e_i^+\rightarrow f_k
\bar{f}_l)=}}&\nonumber\\ &&\!\!\displaystyle{
\frac{e^4s^2C_{\rm{fac}}}{4}\left\{
\left(1+\cos\theta\right)^2
\left[|f^s_{LR}|^2+|f^s_{RL}|^2+|f^t_{LR}|^2+|f^t_{RL}|^2 +2{\rm
Re}(f^{s*}_{LR}f^t_{LR})+2{\rm
Re}(f^{s*}_{RL}f^t_{RL})\right]\right.}\nonumber\\
&&\displaystyle{\left.+(1-\cos\theta)^2\left[|f^s_{LL}|^2+|f^s_{RR}|^2\right]
+4\left[|f^t_{LL}|^2+|f^t_{RR}|^2\right]\rule{0mm}{5mm}\right\}},
\end{eqnarray}
  where the coefficients are given by 
\begin{eqnarray}
f^s_{LL} &=& \delta_{lk}\left\{\frac1{s}
		+\frac{L_eR_e}{s-\mz^2+i\Gambr_Z\mz}\right\}
		+\sum_{j=1,3}\frac{\lam_{jik}\lam_{jil}}
			{2e^2\left(t-M^2_{\nut_j}\right)},\\
f^s_{LR} &=& \delta_{lk}\left\{\frac1{s}
		+\frac{L^2_e}{s-\mz^2+i\Gambr_Z\mz}\right\},\\
f^s_{RL} &=& \delta_{lk}\left\{\frac1{s}
		+\frac{R^2_e}{s-\mz^2+i\Gambr_Z\mz}\right\},\\
f^s_{RR} &=& \delta_{lk}\left\{\frac1{s}
		+\frac{L_eR_e}{s-\mz^2+i\Gambr_Z\mz}\right\}
		+\sum_{j=1,3}\frac{\lam_{jki}\lam_{jli}}
			{2e^2\left(t-M^2_{\nut_j}\right)},\\
f^t_{LL} &=& \delta_{il}\delta_{lk}\left\{\frac1{t}
		+\frac{L_eR_e}{t-\mz^2}\right\}
		+\sum_{j=1,3}\frac{\lam_{jii}\lam_{jkl}}
			{2e^2\left(s-M^2_{\nut_j}+i\Gambr_{\nut_j}M_{\nut_j}\right)},\\
f^t_{LR} &=& \delta_{il}\delta_{lk}\left\{\frac1{t}
		+\frac{{L_e}^2 }{t-\mz^2}\right\},\\
f^t_{RL} &=& \delta_{il}\delta_{lk}\left\{\frac1{t}
		+\frac{{R_e}^2 }{t-\mz^2}\right\},\\
f^t_{RR} &=& \delta_{il}\delta_{lk}\left\{\frac1{t}
		+\frac{L_eR_e}{t-\mz^2}\right\}
		+\sum_{j=1,3}\frac{\lam_{jii}\lam_{jlk}}
			{2e^2\left(s-M^2_{\nut_j}+i\Gambr_{\nut_j}M_{\nut_j}\right)}.
\end{eqnarray}
  for $e_i^-e_i^+\rightarrow e^-_k e^+_l$ and by
\begin{eqnarray}
f^s_{LL} &=& \delta_{lk}\left\{-\frac{e_d}{s}
		+\frac{L_eR_d}{s-\mz^2+i\Gambr_Z\mz}\right\}
		+\sum_{j=1,3}\sum_{\al=1,2}\frac{\lam'_{ijk}\lam'_{ijl}}
			{2e^2\left(t-M^2_{\upt_{j\al}}\right)},\\
f^s_{LR} &=& \delta_{lk}\left\{-\frac{e_d}{s}
		+\frac{L_eL_d}{s-\mz^2+i\Gambr_Z\mz}\right\},\\
f^s_{RL} &=& \delta_{lk}\left\{-\frac{e_d}{s}
		+\frac{R_eR_d}{s-\mz^2+i\Gambr_Z\mz}\right\},\\
f^s_{RR} &=& \delta_{lk}\left\{-\frac{e_d}{s}
		+\frac{R_eL_d}{s-\mz^2+i\Gambr_Z\mz}\right\},\\
f^t_{LL} &=&\sum_{j=1,3}\frac{\lam_{jii}\lam'_{jkl}}
			{2e^2\left(s-M^2_{\nut_j}+i\Gambr_{\nut_j}M_{\nut_j}\right)},\\
f^t_{LR} &=&0,\\
f^t_{RL} &=&0,\\
f^t_{RR} &=&\sum_{j=1,3}\frac{\lam_{jii}\lam'_{jlk}}
			{2e^2\left(s-M^2_{\nut_j}+i\Gambr_{\nut_j}M_{\nut_j}\right)},
\end{eqnarray}
  for $e^-_ie^+_i\to d_k \bar{d}_l$.
  The colour factor $C_{\rm{fac}}$ is 1 for charged lepton production and
$N_C$ for quark production. The left/right charges of the fermions are
given in Sect.~\ref{notation}.
  This formula corrects the sign of the interference term given in
\cite{Kalinowski:1997bc}, which is necessary both to reproduce their
numerical results and to obtain the correct \SM\ Bhabha cross section
formula in the limit that the R-parity violating couplings are zero.

 \subsection{Decay matrix elements}
  The decay matrix elements are implemented for both R-parity conserving
and R-parity violating three-body decays of the gluino and the gauginos.
  The R-parity conserving decays are performed using the helicity
amplitude expressions given in Ref.~\cite{Richardson:2001df}.
  The R-parity violating formulae are as given in
Ref.~\cite{Dreiner:1999qz}.

  Spin correlations in the distribution of cascade decay products are
implemented in all decay processes, as well as the R-parity conserving
$2\to2$ \SY\ production processes. The implementation is as described in
Ref.~\cite{Richardson:2001df}.


 \section{Conclusions}
 \label{conclusions}

  The \MC\ event generator \HW\ has been extended to include the
simulation of supersymmetry. We have summarised its salient features. We
have defined the parameter conventions and listed the matrix elements for
all original calculations.



 \subsection*{Acknowledgements}
  We thank the co-authors of HERWIG for fruitful collaboration, and the
members of the Cavendish ATLAS--HERWIG working group for stimulating
discussions and for testing part of our code. P.R. would like to thank
H.~Dreiner for many useful discussions on the implementation of R-parity
violating processes. S.M.\ thanks M.L.~Mangano, H.~Dreiner and M.~Kr\"amer
for many useful discussions and the latter in particular for making
available an independent implementation of the leading order matrix
elements for gaugino production.


 \vfill\pagebreak

 \appendix
 \section{Process codes}
  We summarize the process codes {\tt IPROC} for \MSSM\ hard scattering
processes implemented in \HW.

\begin{center}
\small
\begin{tabular}{|c|l|}
\hline
 {\tt IPROC} &  Process \\
\hline
   700-99  & R-parity conserving \SY\ processes \\
   700     & $\l^+ \l^- \to$~2-sparticle processes (sum of 710--760)\\
   710     & $\l^+ \l^- \to$~neutralino pairs (all neutralinos) \\
706+4{\tt IN1}+{\tt IN2} &$\l^+ \l^- \to \gaugino^0_{\mbox{\scriptsize 
IN1}}
                         \gaugino^0_{\mbox{\scriptsize IN2}}$
                      ({\tt IN1,2}=neutralino mass eigenstate)\\
   730     & $\l^+ \l^- \to$~chargino pairs (all charginos) \\
728+2{\tt IC1}+{\tt IC2} &$\l^+ \l^- \to \gaugino^+_{\mbox{\scriptsize 
IC1}}
                         \gaugino^-_{\mbox{\scriptsize IC2}}$
                      ({\tt IC1,2}=chargino mass eigenstate) \\
   740     & $\l^+ \l^- \to$~slepton pairs (all flavours) \\
   736+5\IL& $\l^+ \l^- \to \slepton_{L,R} \slepton_{L,R}^*$
             ($\IL=1,2,3$ for 
$\slepton=\tilde{e},\tilde{\mu},\tilde{\tau}$) \\
   737+5\IL& $\l^+ \l^- \to \slepton_{L} \slepton_{L}^*$ (\IL\ as above) 
\\
   738+5\IL& $\l^+ \l^- \to \slepton_{L} \slepton_{R}^*$ (\IL\ as above)\\
   739+5\IL& $\l^+ \l^- \to \slepton_{R} \slepton_{R}^*$ (\IL\ as above)\\
   740+5\IL& $\l^+ \l^- \to \snu_{L} \snu_{L}^*$
             ($\IL=1,2,3$ for $\snu_e, \snu_\mu, \snu_\tau$) \\
   760      & $\l^+ \l^- \to$~squark pairs (all flavours) \\
   757+4\IQ & $\l^+ \l^- \to \squark_{L,R} \squark^*_{L,R}$
             ($\IQ=1...6$ for $\squark=\tilde{d}...\tilde{t}$)\\
   758+4\IQ & $\l^+ \l^- \to \squark_{L} \squark^*_{L}$
                (\IQ\ as above)\\
   759+4\IQ & $\l^+ \l^- \to \squark_{L} \squark^*_{R}$
                (\IQ\ as above)\\
   760+4\IQ & $\l^+ \l^- \to \squark_{R} \squark^*_{R}$
                (\IQ\ as above)\\
\hline
 %
   800-99  & R-parity violating \SY\ processes \\
   800     & Single sparticle production, sum of 810--840 \\
   810     & $\l^+ \l^- \to \gaugino^0 \nu_i$, (all neutralinos)\\
   810+\IN & $\l^+ \l^- \to \gaugino^0_{\mbox{\scriptsize IN}} \nu_i$,
             (\IN=neutralino mass state)\\
   820     & $\l^+ \l^- \to \gaugino^- e^+_i$ (all charginos) \\
   820+\IC & $\l^+ \l^- \to \gaugino^-_{\mbox{\scriptsize IC}} e^+_i$,
             (\IC=chargino mass state) \\
   830     & $\l^+ \l^- \to \snu_i Z^0$ and
             $\l^+ \l^- \to \slepton^+_i W^-$  \\ 
   840     & $\l^+ \l^- \to \snu_i h^0/H^0/A^0$ and
             $\l^+ \l^- \to \slepton^+_i H^-$  \\
   850     & $\l^+ \l^- \to \snu_i \gamma$ \\
   860     & Sum of 870 and 880 \\
   870     & $\l^+ \l^- \to \l^+ \l^-$, via LLE only \\
   867+3{\tt IL1}+{\tt IL2} &
$\l^+ \l^- \to \l^+_{\mbox{\scriptsize IL1}} \l^-_{\mbox{\scriptsize 
IL2}}$
          ({\tt IL1,2}=1,2,3 for $e,\mu,\tau$) \\
   880     & $\l^+ \l^- \to \bar d  d$, via LLE and LQD \\
   877+3{\tt IQ1}+{\tt IQ2} &
$\l^+ \l^- \to d_{\mbox{\scriptsize IL1}} \bar d_{\mbox{\scriptsize IL2}}$
          ({\tt IQ1,2}=1,2,3 for $d,s,b$) \\
\hline
 %
    910    &      $\ell^+ \ell^- \to \nu_e \bar\nu_e h^0 + e^+ e^- h^0$\\
    920    &      $\ell^+ \ell^- \to \nu_e \bar\nu_e H^0 + e^+ e^- H^0$\\
\hline
    960    &      $\ell^+ \ell^- \to Z^0 h^0$\\   
    970    &      $\ell^+ \ell^- \to Z^0 H^0$\\
\hline
    955    &      $\ell^+ \ell^- \to H^+ H^-$\\
    965    &      $\ell^+ \ell^- \to A^0 h^0$\\
    965    &      $\ell^+ \ell^- \to A^0 H^0$\\
\hline
\end{tabular}
\\\vspace{3mm}{{\bf Table 6:} The \MSSM\ hard scattering processes 
implemented in \HW\ (continued on following pages).}
\end{center}

 \vfill\pagebreak

\begin{center}
\small
\begin{tabular}{|c|l|}
\hline
{\tt IPROC} & Process\\ 
\hline
   3000-999& R-parity conserving \SY\ processes\\
   3000    & 2-parton $\to$ 2-sparticle processes (sum of those below)\\
   3010    & 2-parton $\to$ 2-sparton processes \\
   3020    & 2-parton $\to$ 2-gaugino processes \\
   3030    & 2-parton $\to$ 2-slepton processes \\ 
\hline
 %
   3100+\ISQ& $gg/q\qbar\to {\tilde q}{\tilde q}^{'*} {H^\pm}$
       ({\it cf.}\ Ref.~\cite{HERWIG6} for definition of \ISQ)\\
\hline
   3200+\ISQ& $gg/q\qbar\to {\tilde q}{\tilde q}^{'*} {h,H,A}$ ('') \\
\hline
   3310,3315    & $q\qbar' \to W^\pm h^0,H^\pm h^0$  \\
   3320,3325    & $q\qbar' \to W^\pm H^0,H^\pm H^0$  \\
      3335      & $q\qbar' \to H^\pm A^0$ \\
      3350      & $q\qbar  \to W^\pm H^\mp$  \\
      3355      & $q\qbar  \to H^\pm H^\mp $  \\
   3360,3365    & $q\qbar  \to Z^0 h^0,A^0 h^0$ \\
   3370,3375    & $q\qbar  \to Z^0 H^0,A^0 H^0$ \\
\hline
   3410      & $bg \to b~h^0$ + ch.\ conj.\\
   3420      & $bg \to b~H^0$ + ch.\ conj.\\
   3430      & $bg \to b~A^0$ + ch.\ conj.\\
   3450      & $bg \to t~H^-$ + ch.\ conj.\\
\hline
   3500     & $b q \to b q' H^\pm$ + ch.\ conj. \\
\hline
   3610& $q\qbar/gg \to h^0$  \\
   3620& $q\qbar/gg \to H^0$  \\
   3630& $q\qbar/gg \to A^0$ \\
\hline
   3710& $q\qbar \to q'\qbar' h^0$  \\
   3720& $q\qbar \to q'\qbar' H^0$  \\
\hline
   3810+\IQ& $gg+q\qbar\to Q\Qbar h^0$ (\IQ\ for $Q$ flavour) \\
   3820+\IQ& $gg+q\qbar\to Q\Qbar H^0$ ('') \\
   3830+\IQ& $gg+q\qbar\to Q\Qbar A^0$ ('') \\
3839~~~~~~~& $gg+q\qbar\to b\bar t H^+$ + ch. conjg. \\
   3840+\IQ& $gg       \to Q\Qbar h^0$ (\IQ\ as above) \\
   3850+\IQ& $gg       \to Q\Qbar H^0$ ('') \\
   3860+\IQ& $gg       \to Q\Qbar A^0$ ('') \\
3869~~~~~~~& $gg       \to b\bar t H^+$ + ch. conjg.  \\
   3870+\IQ& $q\qbar   \to Q\Qbar h^0$ (\IQ\ as above) \\
   3880+\IQ& $q\qbar   \to Q\Qbar H^0$ ('') \\
   3890+\IQ& $q\qbar   \to Q\Qbar A^0$ ('') \\
3899~~~~~~~& $q\qbar   \to b\bar t H^+$ + ch. conjg. \\
\hline
 %
\end{tabular}
\\\vspace{3mm}{Table~6 continued.}\\
\end{center}

 \vfill\pagebreak

\begin{center}
\small
\begin{tabular}{|c|l|}
\hline
 {\tt IPROC} & Process\\ 
\hline
   4000-99   &  R-parity violating \SY\ processes via LQD\\
   4000      & single sparticle production, sum of 4010--4050 \\
   4010      & $\ubar_j d_k \to \gaugino^0 l^-_i$,
               $\dbar_j d_k \to \gaugino^0 \nu_i$ (all neutralinos)\\
  4010+\IN  & $\ubar_j d_k \to \gaugino^0_{\mbox{\scriptsize IN}} l^-_i$,
               $\dbar_j d_k \to \gaugino^0_{\mbox{\scriptsize IN}} \nu_i$
(\IN=neutralino mass state)\\
   4020      & $\ubar_j d_k \to \gaugino^- \nu_i$,
               $\dbar_j d_k \to \gaugino^- e^+_i$ (all charginos) \\
   4020+\IC & $\ubar_j d_k \to \gaugino^-_{\mbox{\scriptsize IC}} \nu_i$,  
               $\dbar_j d_k \to \gaugino^-_{\mbox{\scriptsize IC}} e^+_i$ 
(\IC=chargino mass state)\\
   4040      & $u_j \dbar_k \to \tilde{\tau}^+_i Z^0$,
               $u_j \dbar_k \to \snu_i W^+$ and
               $d_j \dbar_k \to \slepton^+_i W^-$  \\
   4050      & $u_j \dbar_k \to \slepton^+_i h^0/H^0/A^0$,
               $u_j \dbar_k \to \snu_i H^+$ and    
               $d_j \dbar_k \to \slepton^+_i H^-$  \\
   4060      & Sum of 4070 and 4080 \\
   4070      & $\ubar_j d_k \to \ubar_l d_m $ and
               $\dbar_j d_k \to \dbar_l d_m $, via LQD only \\
   4080      & $\ubar_j d_k \to \nu_j l^-_k $ and
               $\dbar_j d_k \to l^+_j l^-_k $, via LQD and LLE \\
\hline
   4100-99 & R-parity violating \SY\ processes via UDD\\
   4100    & single sparticle production, sum of 4110--4150\\
   4110      & $u_i d_j \to \gaugino^0 \dbar_k$,
               $d_j d_k \to \gaugino^0 \bar{u_i}$ (all neutralinos)\\
  4110 +\IN  & $u_i d_j \to \gaugino^0_{\mbox{\scriptsize IN}} \dbar_k$,
               $d_j d_k \to \gaugino^0_{\mbox{\scriptsize IN}}
 \bar{u_i}$(\IN\ as above)\\
  4120       & $u_i d_j \to \gaugino^+ \ubar_k$,
               $d_j d_k \to \gaugino^- \bar{d_i}$  (all charginos) \\
  4120 +\IC  & $u_i d_j \to \gaugino^+_{\mbox{\scriptsize IC}} \ubar_k$,
               $d_j d_k \to \gaugino^-_{\mbox{\scriptsize IC}}
 \bar{d_i}$  (\IC\ as above) \\
  4130       & $u_i d_j \to \gluino \dbar_k$,
               $d_j d_k \to \gluino \bar{u_i}$ \\
  4140       & $u_i d_j \to \tilde{b}^*_1 Z^0$,  
$d_j d_k \to \tilde{t}^*_1 Z^0$,
               $u_i d_j \to \tilde{t}^*_i W^+$
and $d_j d_k \to \tilde{b}^*_i W^-$  \\
  4150       & $u_i d_j \to \tilde{d}^*_{k1} h^0/H^0/A^0$,
               $d_j d_k \to \tilde{u}^*_{i1} h^0/H^0/A^0$,
               $u_i d_j \to \tilde{u}^*_{k\alpha} H^+$,
               $d_j d_k \to \tilde{d}^*_{i\alpha} H^-$  \\
  4160       & $u_i d_j \to u_l d_m$, $d_j d_k \to d_l d_m$ via UDD. \\
\hline
\end{tabular}
\\\vspace{3mm}{Table~6 concluded.}\\
\end{center}  


\end{document}